\documentclass[useAMS,usenatbib]{mn2e}
\usepackage{graphicx,graphics, amsmath,lscape,subfigure,times,color}
\title{Environments of Extended Radio Sources in the ATLBS Survey}
\author[K. Thorat et al.] {K. Thorat$^{1,2}$,L. Saripalli$^1$, R. Subrahmanyan$^{1,3}$ \\
$^{1}$Raman Research Institute, C. V. Raman Avenue, Sadashivanagar, Bangalore 560080, India \\
$^{2}$Joint Astronomy Programme, Indian Institute of Science, Bangalore 560012, India \\
$^{3}$National Radio Astronomy Observatory, Socorro, NM 87801, USA}

\begin{document}
\date{}
\maketitle
\label{firstpage}
\begin{abstract}
We present a study of the environments of extended radio sources in the Australia Telescope Low Brightness Survey (ATLBS). The radio sources were selected from the Extended Source Sample (ATLBS-ESS), which is a well defined sample containing the most extended of radio sources in the ATLBS sky survey regions. The environments were analyzed using 4-m CTIO Blanco telescope observations carried out for ATLBS fields in the SDSS ${\rm r}^{\prime}$ band. We have estimated the properties of the environments using smoothed density maps derived from galaxy catalogs constructed using these optical imaging data. The angular distribution of galaxy density relative to the axes of the radio sources has been quantified by defining anisotropy parameters that are estimated using a new method presented here. Examining the anisotropy parameters for a sub-sample of extended double radio sources that includes all sources with pronounced asymmetry in lobe extents, we find good evidence for environmental anisotropy being the dominant cause for lobe asymmetry in that higher galaxy density 
occurs almost always on the side of the shorter lobe, and this validates the usefulness of the method proposed and adopted here. The environmental anisotropy parameters have been used to examine and compare the environments of FRI and FRII radio sources in two redshift regimes ($z<0.5$ and $z>0.5$).  Wide-angle tail sources and Head-tail sources lie in the most overdense environments. The Head-tail source environments (for the HT sources in our sample) display dipolar anisotropy in that higher galaxy density appears to lie in the direction of the tails. Excluding the Head-tail and Wide-angle tail sources, subsamples of FRI and FRII sources from the ATLBS survey appear to lie in similar moderately overdense environments, with no evidence for redshift evolution in the regimes studied herein.   
\end{abstract}
 

\section{Introduction}
\label{Section_introduction}

The Australia Telescope low-brightness survey (ATLBS; \citet{SESS10}) is a radio continuum survey at 1.4~GHz
of a moderately large region of 8.4~square degrees of the southern sky.  The ATLBS survey has
imaged radio sources with excellent  surface brightness sensitivity and hence constitutes a useful resource for studies of structural types.  High resolution radio images of the survey regions are presented in \citet{TSSE12} along with a discussion of the source counts. \newline

A subset of extended radio sources detected in the survey constitutes the ATLBS extended source 
sample (ATLBS-ESS; \citet{SST11}); it contains radio galaxies
observed to have the largest angular size and includes all sources with angular size $>30^{\prime\prime}$. 
The examination of ATLBS-ESS source structures has yielded subsamples of restarted radio galaxy candidates, 
$z>0.5$ low-power radio galaxies, giant radio galaxies and other morphological types. The variety of radio structures detected and their relative abundance has been used to infer the life cycles of radio sources. \newline

Obviously, the gas environments in which radio sources reside and evolve ought to have a substantial influence on the structures that form; this expectation has been vindicated in many case studies where the radio structures have been compared with the X-ray gas environments (\citet{BRC11,BVF93}).    
Numerous studies of the optical environments of radio galaxies have been carried out previously 
\citep{LS79,YG1984,HL91,Z97}. The motivation behind these studies has been to examine differences 
between different classes of radio sources, the evolution of environments with cosmic epoch as well as the 
possibility of identifying clusters/groups of galaxies using radio sources as a tracer \citep{WB11}. 
\newline

Several studies have found that the environments of FRI/FRII sources are different, and have a redshift dependence. 
Although FRI sources are found in richer environments,
FRII sources at low redshifts are mostly observed to be hosted by field galaxies, 
where as at relatively higher redshifts ($z > 0.5$) the FRII environments appear to be richer \citep{HDGM11,BLMR03,OBCV08}. There have been fewer studies that relate the richness of the environments and morphological asymmetries of radio galaxies. Two investigations by \citet{sub08} and \citet{saf09} are noteworthy in this regard where the radio structures of two giant radio galaxies were examined in the context of the large-scale galaxy distributions in their vicinity (also see \citet{che12} and references therein). The study was also used to infer properties of the ambient thermal gas medium in which the structures evolved. Clear correlations between structural asymmetries and associated extended emission-line gas was also found for radio galaxies that have relatively smaller sizes of a few hundred kpc \citep{mcC91}. \newline

Examining the environments of radio galaxy hosts has been one of the primary aims of the ATLBS survey. Towards this goal as well as to obtain properties of the host galaxies 
multi-band optical observations were carried out. In this paper we 
report on the optical observations and use this resource in an attempt to characterize
the environments of the ATLBS-ESS radio sources. \newline

The outline of the paper is as follows. In the next section, we describe the sample definition and selection process. In Section ~\ref{Section_Obs_data_red} we describe the optical observations and data reduction. In Section~\ref{Section_Photometry} we describe the photometry. In Section ~\ref{Section_Red_Mag_Rel}, we derive the redshift-magnitude relation that we subsequently use to estimate redshifts for those sources in the sample that lack spectroscopic redshift measurements. Section~\ref{Section_Quant_gal_env} presents a description of the method adopted to quantify the environmental richness and spatial distribution relative to the radio axes. The last section is a presentation and discussion of the results of the study. We have used a LCDM cosmology with $H_{o}=71$~km~s$^{-1}$~Mpc$^{-1}$ and $\Omega_{\Lambda}=0.73$. \newline

\section{Sample Definition and Selection}
\label{Section_Sample_Def_Sel}
We have chosen to restrict our study of the environments of ATLBS sources to those that are extended and 
hence to a subset of the ATLBS-ESS sources. Briefly, the ATLBS-ESS subsample consists of 119 radio sources that have angular size exceeding $0.5'$. We have omitted sources where no magnitude or redshift information is available. Additionally, since radio galaxies at high redshifts may suffer from greater incompleteness in the detection of neighboring galaxies, we have imposed a redshift cut, choosing only sources below redshift of z $= 1$.  We also reject those sources which are near the edges of the optical images so that the environmental information is not truncated in sky projection: in practice we have rejected sources within a linear distance of $0.5$~Mpc of the image edge (for more discussion on this see Section~\ref{subsection_implementation}).  \newline

\begin{figure*}
\subfigure[]{
\includegraphics[height=3in,width=3.2in]{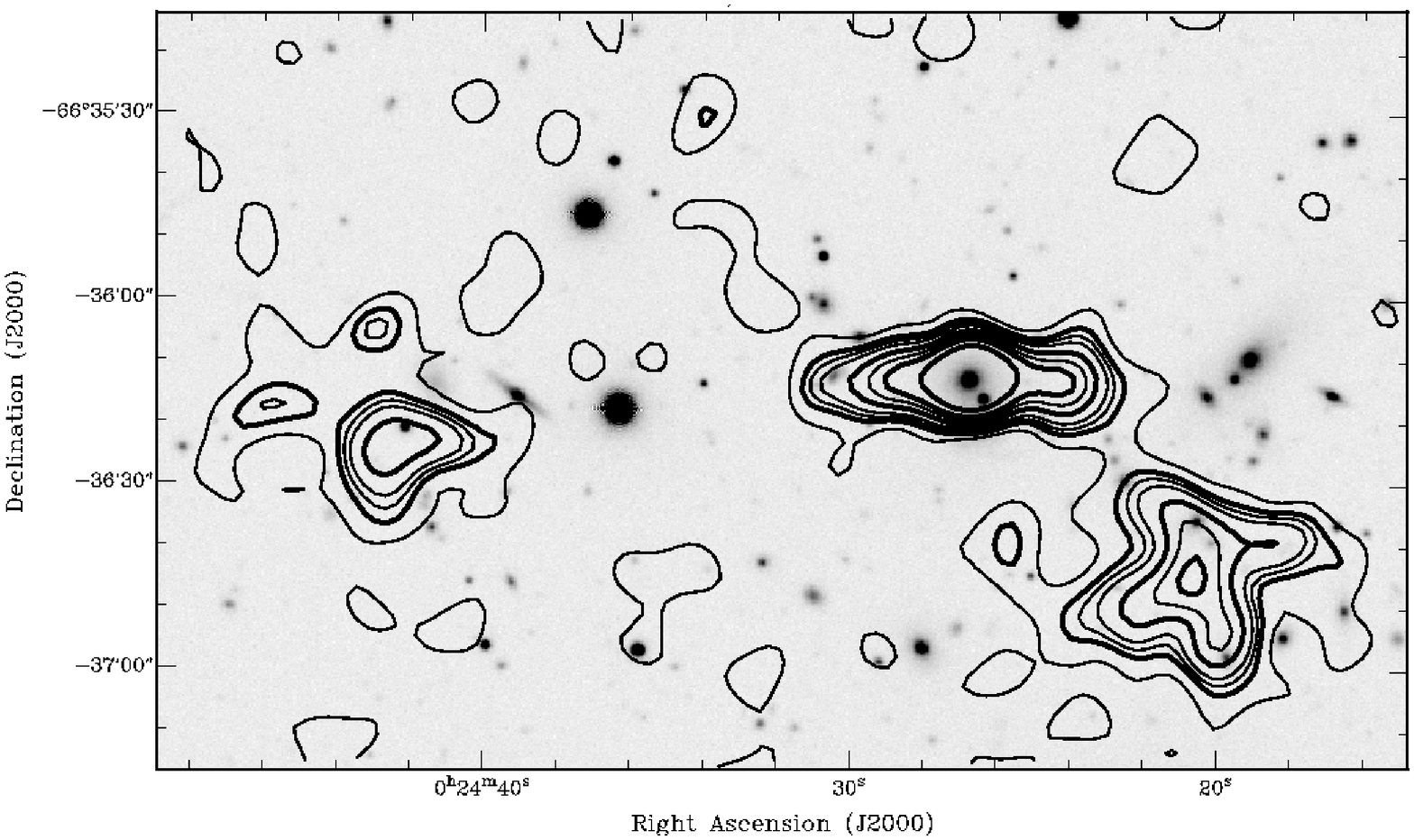}
\label{fig1a}
}
\subfigure[]{
\includegraphics[height=3in,width=2.8in]{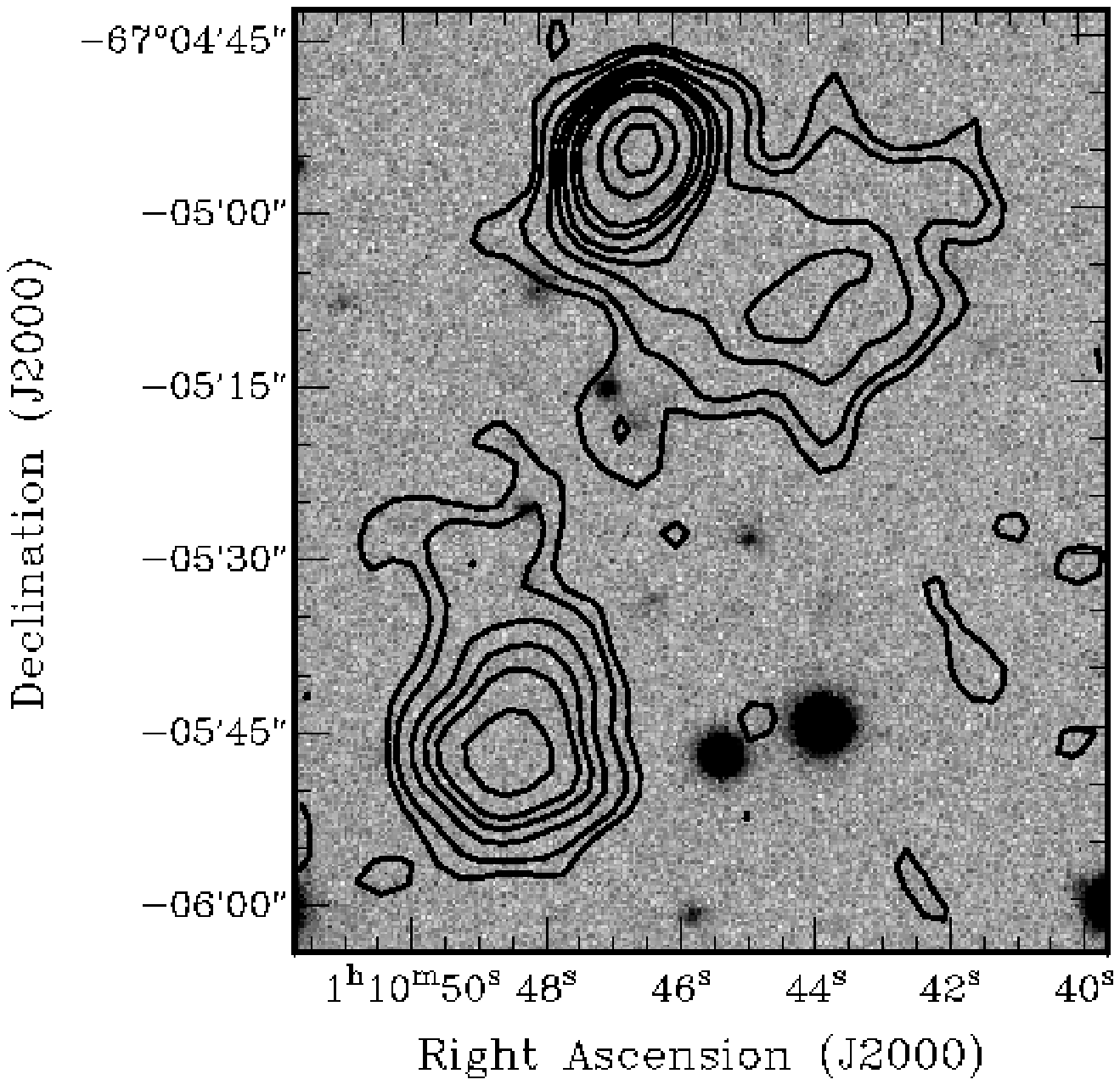}
\label{fig1b}
}\\
\subfigure[]{
\includegraphics[height=3in,width=3in]{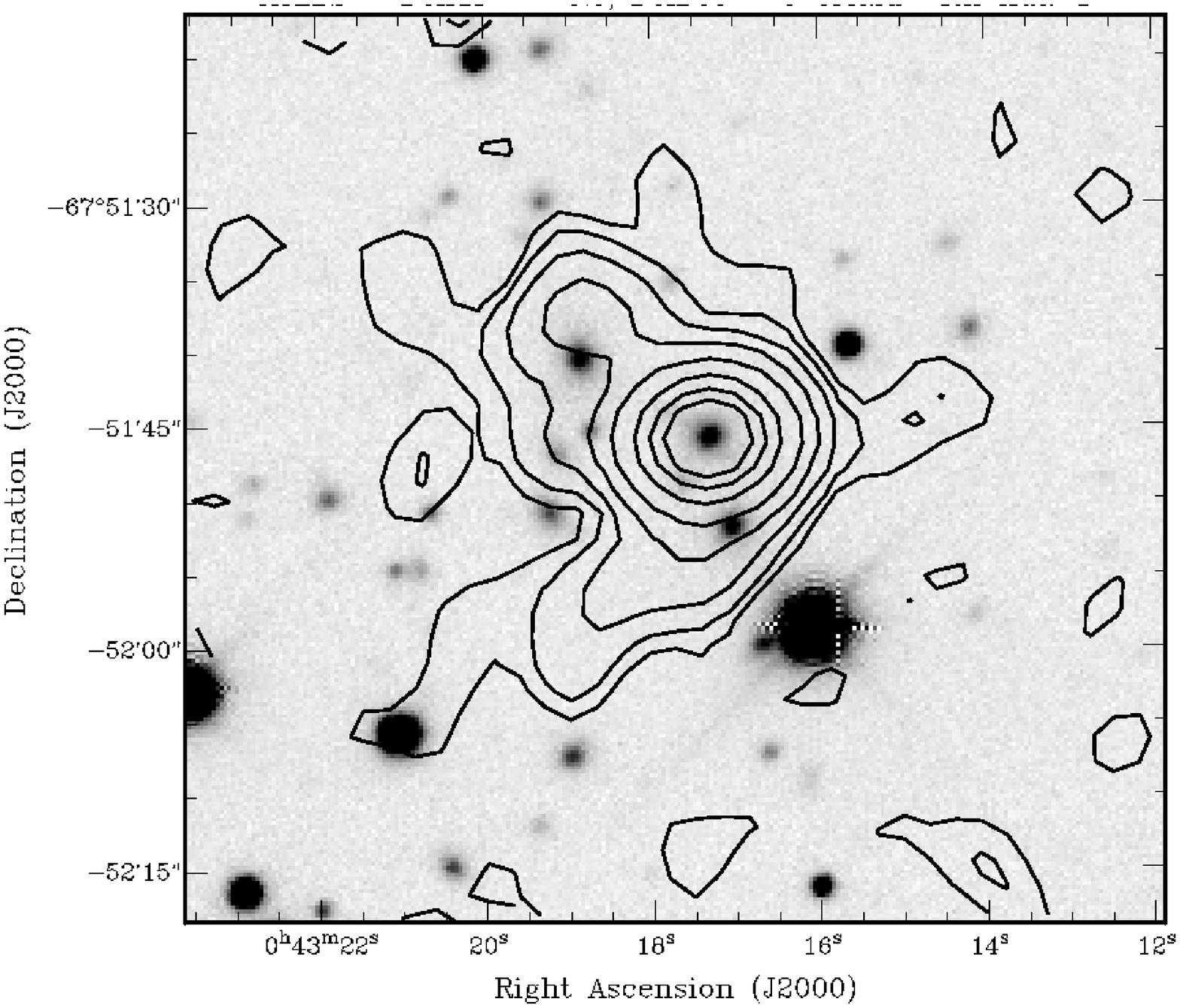}
\label{fig1c}
}
\subfigure[]{
\includegraphics[height=3in,width=3in]{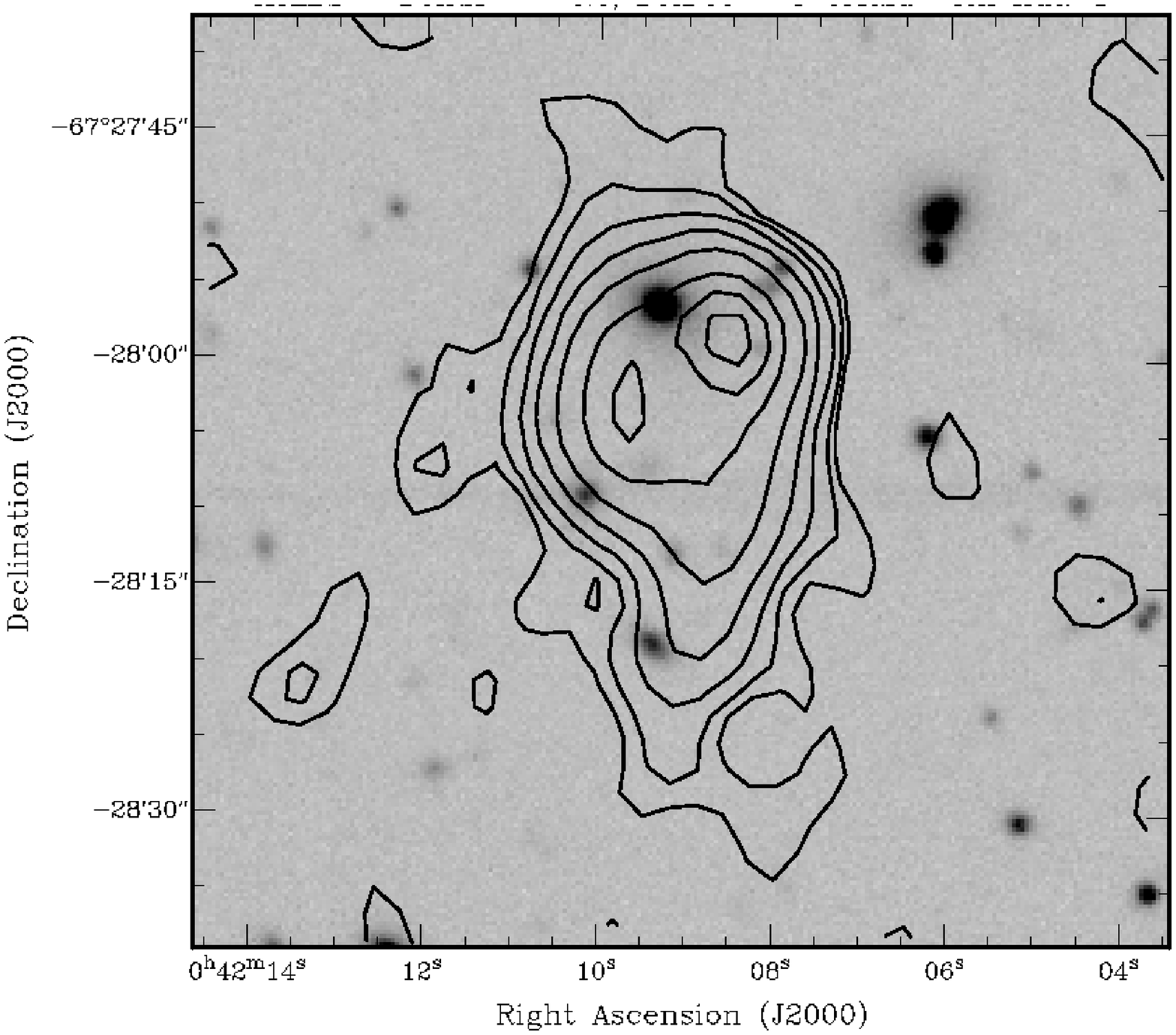}
\label{fig1d}
}
\caption{(a) The radio source J0024.4$-$6636, a FRI source. The source is also an asymmetric source. The radio contours are at 1, 2, 2.5, 3, 4, 5, 6, 8 $\times 10^{-4}$ Jy~beam\textsuperscript{-1} (b) The radio source J0110.7$-$6705, a FRII source. The  radio contours are at 1, 2, 4, 6, 8, 12, 16, 32, 48 $\times 10^{-4}$ Jy~beam\textsuperscript{-1}. (c) The radio source J0043.2$-$6751, a WAT source. The radio contours are at 1, 2, 4, 8, 16, 32, 48, 64 $\times 10^{-4}$ Jy~beam\textsuperscript{-1} (d) The radio source J0042.1$-$6728, a HT source. The radio contours are at 1, 2, 4, 8, 16, 32, 48, 64 $\times 10^{-4}$ Jy~beam\textsuperscript{-1}. The grayscale in each of the above is optical taken from SDSS ${\rm r}^{\prime}$ band images, described in this paper. The examples of the sources given here show the same sources and follow the same contours as Figs. 1.5 (right panel), 1.117, 1.48 and 1.47 from \citet{SST11}.}
\label{lab_fig1}
\end{figure*}

The final sample of sources chosen for the environment study has 43 sources (Fig.~\ref{lab_fig1} presents images of sources of different morphologies from the selected sample). In Table~\ref{od_sample} we present this source list. For these most extended of ATLBS radio sources, with good structural information and hence well classified morphology, we attempt to examine the environments in relation to different source classifications and radio structure. \newline

\section{Observations and data reduction}
\label{Section_Obs_data_red}

The ATLBS survey covers two adjacent regions in the southern sky, which we refer to as ATLBS regions 
A and B.  These are centered at RA: $00^{h}\ 35^{m}\ 00^{s}$, DEC: $-67^{\circ}\ 00^{'}\ 00^{''}$ and 
RA: $00^{h}\ 59^{m}\ 17^{s}$, DEC: $-67^{\circ}\ 00^{'}\ 00^{''}$ (J2000 epoch) respectively. The radio observations 
were carried out using the Australia Telescope Compact Array (ATCA). 
Details of the radio observations and 
imaging have been presented in \citet{SESS10} and \citet{TSSE12}. \newline

Followup optical observations were carried out with the 4-meter Blanco Telescope at Cerro-Tololo Inter-American
Observatory (CTIO), Chile. The observations were carried out in SDSS ${\rm r}^{\prime}$, ${\rm g}^{\prime}$ and 
${\rm z}^{\prime}$ bands, using the MOSAIC II imager. 
Each optical image covers $37' \times 37'$ area in the sky. The MOSAIC II imager covers each image with $2$ rows 
of $8$ CCDs, each of them $2048$ pixels wide and $4096$ pixels long, giving a scale of $0.27''$ per pixel. 
In this work we have used only the ${\rm r}^{\prime}$ band images. \newline

The observations in ${\rm r}^{\prime}$ band were carried out over a complete night and part of the subsequent night. 
The observations were designed so as to cover each of regions A and B with $14$ telescope pointings that 
tiled the individual regions with some overlap.  Each pointing position was observed as 5 consecutive integrations that
were made with the telescope pointing slightly dithered from the nominal pointing position: one integration was with the 
telescope at the pointing center and the remaining four were made with about arcmin offsets towards N, S, E and W.  The multiple 
dithered exposures were made to exclude artifacts associated with CCD errors as well as cosmic rays.
Each pointing was observed for $5 \times 100$ seconds.   The observing night included six exposures 
towards standard stars; calibration data includes bias exposures and flat fields. \newline 

The calibration and construction of images from this data was carried out with the Image 
Reduction and Analysis Facility (IRAF) software. 
We have used the `MSCRED' package of IRAF extensively, which has been written for use with mosaic data. 
Each image was bias and overscan corrected, corrected for cross talk, trimmed and flat fielded in the 
calibration process. The images have an overscan region of 50 pixels and this was used to correct the entire image for 
the mean bias. 
We have in total 20 bias frames that were combined to form a master bias frame, which was used for correction of 
variable bias in the images. The cross talk correction was done using the 
cross talk correction files provided for the MOSAIC II imager by NOAO. We have also identified and generated 
masks that identify erroneous pixels in each of the images. These contain saturated pixels and bleed trails 
along with pixels known to be faulty in the MOSAIC II imager. The saturated pixels were identified with the help of the saturation values recorded in the header. The bleed trails were identified
using a minimum trail length of $20$ pixels and a pixel value one-third of saturation value. We removed the 
bleed trails in the images by excluding these pixels and interpolating over them using surrounding pixels. 
For the purpose of flat fielding, dome flats were combined to make a master 
dome flat. For a more accurate flat-fielding, we have constructed a `super sky-flat' from observed images. 
For doing this, we first created object masks for all the target frames, removed the objects from the images using the masks 
and then combined them into a single image that represents the super sky flat. After flat-fielding using this image the variations in the 
image response are estimated to be at the $2\%$ level. \newline

We found that the world coordinate system (WCS henceforth) attached to the raw images was inaccurate. An accurate WCS is 
essential for correct stacking of the dithers and, more importantly, for using the images to examine the environments of the radio sources that are in the ATCA radio images. Therefore, we used USNO A2.0 R band catalogs of objects to reconstruct a correct WCS for the images. Objects common to the
USNO catalog and the images were chosen and their positions were matched to construct the new WCS. The resultant WCS 
shows an rms difference of $0.3''$ compared to positions of USNO catalog objects. \newline

After correcting the WCS, the images corresponding to the 5-position dithers were stacked. First, each of the dithered images was tangent plane projected using sinc-function interpolation; second, the images were rescaled to have the same background brightness.  
The stacking was carried out using averaging 
with sigma clipping. The stacking procedure removes spurious objects in the dithers, such as satellite trails and 
cosmic ray trails, as these may not appear in different dithers at the same pixel. The stacked and co-added images were used for the analyses presented herein. \newline

\section{Photometry}
\label{Section_Photometry}
We have used IRAF to carry out the initial photometry using exposures on standard stars and derive the photometric parameters for calibration of the target fields.  We observed standard stars from the field of NGC~458~AB\footnote{http://www-star.fnal.gov/Southern\_ugriz/Old/www/NGC\_458-AB.html} \citep{AWA95};  in the ${\rm r}^{\prime}$ band three standard star exposures were made during the first night and one standard star exposure during the second night. \newline

The aim of the photometry is to obtain a relation between the measured instrumental magnitudes and true magnitudes. Since we have target images observed at different airmass, we require to derive the calibration relation as a function of airmass. We fit for a linear measurement equation with two free parameters: an additive term and a coefficient that multiplies the airmass. The relation used is $m_{s}=C_{1}*A + C_{2}$ where $C_{1}$ and $C_{2}$ are the free parameters. We have used 33 standard stars covering a wide range in intensity to fit for the photometric parameters, and used the IRAF task PHOT to derive the
instrumental magnitudes and the PHOTCAL package for the fitting. \newline

For deriving the magnitudes from the optical images we used Source Extractor \citep{BA96} in conjunction with the initial photometry as explained above. For object detection, we select pixels above $2.5\ \sigma$, where $\sigma$ is the standard deviation of the image rms noise. The rms noise is determined from an area of 64 pixel square around the source. In the vicinity of bright sources the background is modeled using the point spread function and the detection threshold is appropriately raised.  We find that for
our data, the Kron-like apertures of Source Extractor is well suited to estimating the instrumental magnitudes. 
This measures the flux in apertures of size $2.5\times r_{\rm eff}$, where $r_{\rm eff}$ is the effective radius given by the first 
moment of intensity distribution. The flux detected in the aperture is background subtracted; for this the background is determined from image pixels in an annulus of width $20$ pixels around the objects. \newline

Source Extractor also identifies objects by estimating the stellarity index in the range $0.0 - 1.0$, where a value of $1.0$ corresponds to a star-like object and $0.0$ is almost certainly a galaxy.
This parameter has been widely used in the literature for the purpose of star-galaxy classification: {\it e.g.} \citet{VDMF09} use it for classification of Wide-field Nearby Galaxy-cluster Survey (WINGS) survey data, preferring to use it with constrains on the stellarity index. In particular, they catalog objects with stellarity index less than $0.2$ as galaxies  and objects greater than $0.8$ as stars. \newline

\begin{figure*}
\includegraphics[width=\textwidth]{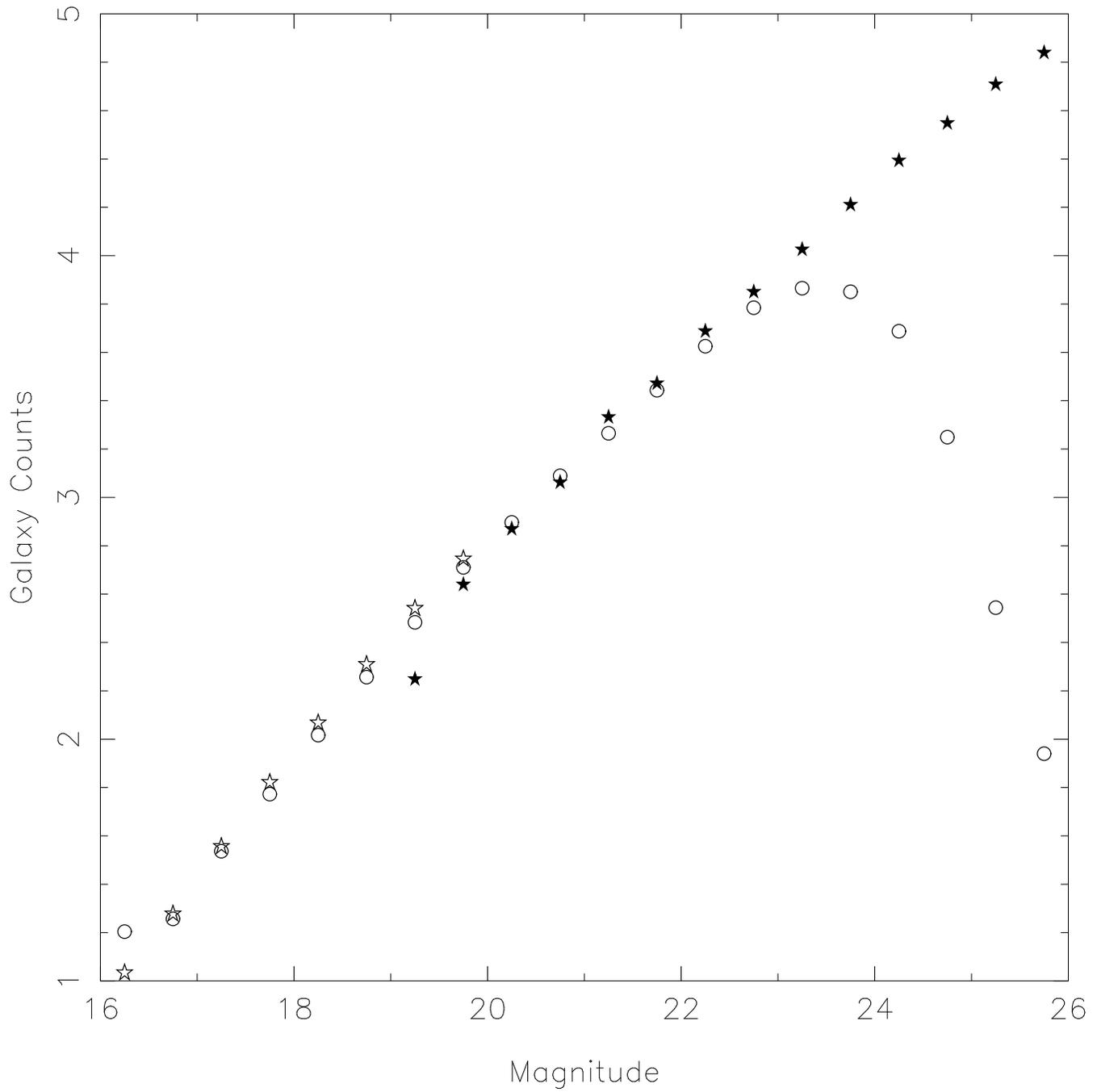}
\caption{The ${\rm r}^{\prime}$ band galaxy counts for the ATLBS survey region (displayed using unfilled circles). For comparison with literature, the counts from \citet{ZHAF09} are shown using filled stars and that of \citet{YFNL01} are shown using unfilled stars.}
\label{galcnt}
\end{figure*}

We have estimated galaxy counts from the object catalog created by Source Extractor.  
Objects with a stellarity index between $0.0$ and $0.4$ were assumed to be `galaxies'. We have chosen 0.4 as a more relaxed upper 
limit for the stellarity index to be more inclusive and to avoid losing galaxies with higher stellarity indices.  
This relaxation admittedly increases the risk of including spurious stars; however, they may be expected only to affect the uncertainty 
in the parameter estimates without causing systematic errors. In Fig.~\ref{galcnt} we compare the derived ATLBS galaxy counts with those 
from the literature. We have compared our galaxy counts with \citet{ZHAF09} as well \citet{YFNL01}. 
While \citet{ZHAF09} present observations of the Extended Groth Strip (EGS) field in u, g and R bands (with their R-band differential counts going deeper than ours), \citet{YFNL01} present galaxy counts in SDSS u$^{\prime}$, g$^{\prime}$, r$^{\prime}$, i$^{\prime}$,  and z$^{\prime}$ filters.  
Our derived counts (complete to 22.75 mag) agree well with the counts presented in both these works. 

\section{Redshift-magnitude relation}
\label{Section_Red_Mag_Rel}
A study of the galaxy environments of ATLBS radio sources requires estimates for their redshifts. Since only a fraction of the host galaxies of the ATLBS radio sources have spectroscopic redshifts, we derive here a redshift-magnitude relation for the host galaxies of ATLBS sources, which we use in subsequent sections that explore the galaxy environments. \newline

The ATLBS-ESS sample \citep{SST11} was selected based on the angular size of the sources. Of this sample, only $19$ radio galaxies have good quality redshift measurements as well as optical magnitudes.  To have additional data for deriving a better fit for the relation, we included 61 ATLBS sources that were relatively compact and hence 
not in the ATLBS-ESS list.  These are ATLBS radio galaxies that have good quality redshifts available from our on-going AAOmega observations (Johnston et al., in preparation). 
We omit quasars while assembling this sub-sample for deriving the magnitude-redshift relation since their optical magnitudes have contributions from the AGN at the centre.  Most of the sub-sample of radio galaxies have redshifts in the range 0.2--0.5. We present the sources utilized to derive the redshift-magnitude relation in Table.~\ref{fit_sample}.\newline

After producing the $r-z$ plot for these sources, we discovered that $20$ of the sources were systematically offset 
from the principal clustering in the $r-z$ plot.  Some of these 
outliers were spirals or showed spiral-like features; the remainder, though ellipticals, showed clear signs of 
disturbed optical morphology.  Excluding these, we had 80 sources in all to estimate the magnitude-redshift relation. 

\begin{figure*}
 \includegraphics[angle=270,width=6in]{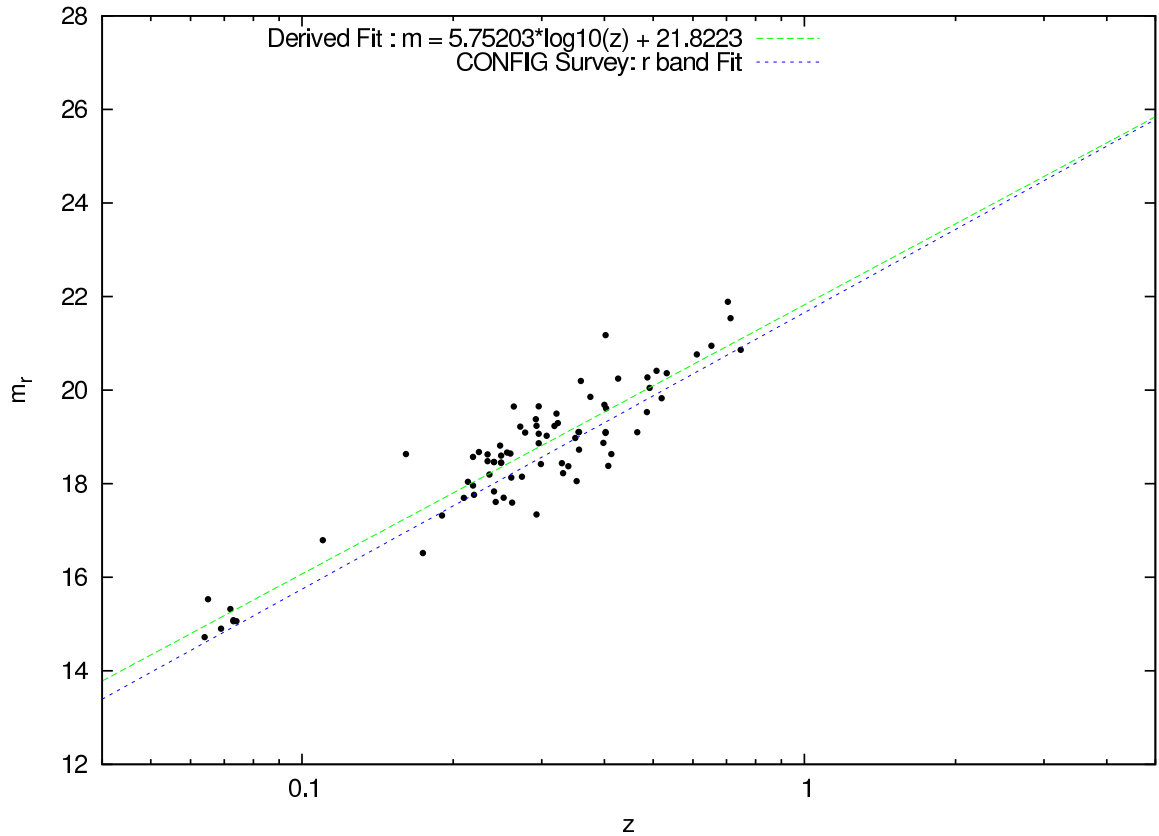}
\caption{The fitted magnitude-redshift relation is shown as lines along with the data points used in deriving the fits; the data points correspond to the compilation of ATLBS radio galaxies with redshift measurements. The average error in z from this relation is  0.09.  For comparison, the relation derived by \citet{GWB11}  (CONFIG Survey) is also shown.}
\label{mz_final_plot}
\end{figure*}

We fit for coefficients of the equation
\begin{equation}
 m_{r} = a\times log_{10}z + b.
\end{equation}  
The fit yielded parameter values $a = 5.752$ and $b = 21.82$, with an rms error of 0.09. These may be compared with corresponding values 
of $a = 5.3$ and $b=21.05$ derived by \citet{ERD97} for $R$ band, and $a=5.917$ and $b = 21.65$ derived by \citet{GWB11}
from the $r'$ band data of the CONFIG survey.   We plot the fitted relation in Fig.~\ref{mz_final_plot} along with the data. \newline

\section{Quantifying the galaxy environments of extended radio sources}
\label{Section_Quant_gal_env}
While there are many methods in the literature for measuring the environmental richness of extragalactic sources (see \citet{Gal08} for a review), most rely on the availability of redshifts. When working with a photometric catalog however, where redshifts are not available for most galaxies, there are fewer methods available for estimating the environmental richness. One of the more common methods used with photometric 
catalogs is the counts-in-a-cell method, which was used by \citet{Ab58} to estimate richness of clusters. In this method the over-density of galaxy counts relative to a mean background density of galaxies is estimated. In determining the environments of radio galaxies, \citet{HL91}, for example, used a variant of this method. However, the latter method has the disadvantage that the galaxies counted in a chosen volume (for example near a radio source) would include contamination from galaxies along the line of sight.  Additionally, the selected volume may not sample the cluster fully, either in spatial or luminosity ({\it i.e.}, magnitude) range.   The above method gives good estimates of the environmental richness in cases where clusters are {\it a~priori} known to exist; however, for the purpose of the blind study we are attempting in the work presented herein, which involves estimating environmental richness for regions with no {\it a~priori} information available, it may not be useful.  Another method that has been used for estimation of environmental richness is that 
using the galaxy-galaxy two-point correlation function \citep{Hard04}, which has been used to detect galaxy clustering around any specified 
point of interest.  The two-point correlation method has the advantage of not being dependent on a particular form for the structure ({\it e.g.} galaxy cluster). \newline  

In the work presented here, we adopt the method of spatial filtering as put forward by \citet{Post96} (P96 henceforth). This method gives the large-scale environment of the source under examination in the form of a map as opposed to quantifying the environmental richness at specific points (e.g. host galaxy location). This enables us to characterize environmental anisotropy on the sky relative to the projected geometry of the source.\newline

\subsection{Description of the Method}
\label{SubSection_Desc_Method}
The spatial filter method, which has been devised for use when only photometric information is available, is as follows. A smoothed galaxy map is created by using a convolving function, which is a composite of two filters: a spatial filter and a magnitude filter. 
The filters are chosen so as to match the density profile of a galaxy cluster and the luminosity function. \newline

The spatial filter is the projected cluster radial profile. The form of the radial filter is given by Eqn.~19 from P96:
\begin{eqnarray}
 P(r/r_{c}) &=& \frac{1}{\sqrt{1+(r/r_{c})^{2}}} \nonumber \\
 && - \frac{1}{\sqrt{1+(r_{co}/r_{c})^{2}}} \qquad \mbox{for} \quad r < r_{co} \nonumber \\
            &=& 0 \qquad \mbox{otherwise},
\end{eqnarray}
where $r_{c}$ is the cluster core radius and $r_{co}$ is the cutoff radius. In literature, a choice of $1 h^{-1}$~Mpc for the cluster cutoff radius has been made, and the core radius has been adopted to be a factor of 10 smaller at $100 h^{-1}$~kpc (P96 and references therein; \citet{Kim02}). The cutoff radius determines the efficiency in the detection of clusters, more than the actual form of the radial filter \citep{Kim02}.  The smoothing is essentially a spatial filter that rejects structures with scale size well below the cutoff radius; therefore, we have used a somewhat smaller cutoff radius of $0.5$~Mpc so that we retain galaxy distribution structures corresponding to relatively poorer clusters.   Following P96, we have used a core radius that is a factor of 10 smaller than the cutoff radius: we use $r_c = 50$~kpc. \newline

The magnitude filter has been chosen to be a Schecter luminosity function with the following form:
\begin{equation}
\phi \propto 0.4 \times ln(10)\\10^{-0.4(m-m_{c})(\alpha + 1)}e^{-10^{-0.4((m-m_{c})}}.
\end{equation}
We have adopted $\alpha = -1.03$ and $m_{c}$, the characteristic magnitude of the luminosity function, to be $-20.6$ (in absolute magnitude units); these are values typical for galaxy clusters \citep{PBRV05}. \newline

The matched filtering essentially creates a smoothed image optimized for the detection of clusters whose properties match the filter characteristics.  Galaxy clustering structure with properties that deviate from the chosen model would be represented in the smoothed image with reduced prominence.  The smoothed image represents the likelihood that a cluster is present at each pixel location and at the redshift of the host galaxy. \newline
 
We may point to a few drawbacks of the method. The form of the smoothing filter is assumed a priori. This means that any over-densities 
in the environments that have a form that deviates substantially from the filter, such as a filamentary structure, will be represented 
with smaller significance. An accurate estimate of the galaxy background (a detailed description of the background is in 
Section~\ref{subsection_implementation} below) is required as a correction to the counts; the background may contain galaxies from 
both cluster as well as non-cluster galaxies and non-uniformity in the distribution may result in errors in the estimate of the 
background. This does not cause problems provided the optical images are large and clusters occupy small sky area; however, in 
smaller images where clusters may be dominant over non-cluster or field galaxies, the erroneous estimate of the background may give 
incorrect results for the inferred structure at the redshift of interest. The optical $r'$ band images used herein have a fairly 
large size ($37' \times 37'$), which obviates the latter concern. \newline

An issue that merits mention is that we have not taken into 
account the redshift dependence of the core and cutoff radius, or the absolute characteristic magnitude of the cluster or their radial profiles; 
the clusters at high redshifts may have substantially different parameters than those we have used. \newline

\subsection{Implementation of the Method}
\label{subsection_implementation}

In this section we follow the nomenclature used by P96. 
We evaluate the output smoothed image with a sampling that is sparse relative to the input image; the output image is evaluated
at its `pixels' as a weighted summation over the input image:
\begin{equation}
 S(i,j) = \sum_{k=1}^{N_{gal}} P[r_{k}(i,j)]L(m_{k}),
\end{equation}
where $L(m_{k})$ corresponds to:
\begin{equation}
 L(m) = \frac{\phi(m-m_{c})}{b(m)} = \frac{\phi(m-m_{c})10^{-0.4(m-m_{c})}}{b(m)}.
\end{equation}
\newline
In the above equation, $b(m)$ is the `background' surface density of galaxies and the factor $10^{-0.4(m-m_{c})}$ has been introduced to keep $L(m)$ integrable (see P96). The sum is evaluated at pixels of the output image denoted by the indices (i, j) and the index k is over all the galaxies in the field ($N_{gal}$ is the total number of galaxies in the field).   $r_{k}$ is the distance from the position of the output pixel to the position of the galaxy with index k, which has an apparent magnitude `$m_{k}$'. \newline

The various terms in the sum $S(i,j)$ are calculated as follows. For a given radio source, for which the environmental richness is to be quantified, a unique smoothed map is created that depends on the source redshift. The characteristic apparent magnitude, the characteristic radius and the cutoff radius for the image corresponding to any radio source depend on its redshift. The characteristic apparent magnitude is determined from the characteristic absolute magnitude using the relation: $M_{c} = m_{c} - DM - K$, where $DM$ is the distance modulus and $K$ is the k-correction appropriate to the redshift of the source. The k-correction for our sources below redshift of $0.7$ used the analytical expressions of \cite{CMZ10} for SDSS ${\rm r}^{\prime}$ band. We have used the luminous red galaxy (LRG) template results for our galaxies, which gives k-correction as a function only of redshift, and yields results similar to that derived by \cite{FSI95} for elliptical galaxies.  Beyond redshift $0.7$ we have used the k-correction given by \cite{Met91} and using galaxy colors given by \cite{FSI95}. The distribution function $b(m)$, which is supposed to be the `background' galaxy counts, is taken to be simply the number counts for galaxies in the field that are fainter than magnitude $m$. This is because we do not know {\it a~priori} which galaxies belong to the background as opposed to clusters.\newline

The normalization for the sum is determined using the following equations (equations 20 and 21 of P96): 
\begin{equation}
 \int_{0}^{\infty} P(r/r_{c})2 \pi r dr = 1
\end{equation}
and
\begin{equation}
\int_{0}^{m_{\rm lim}} \phi(m-m_{c})10^{-0.4(m-m_{c})}dm = 1.
\end{equation}
Here the radial integration is truncated at the cutoff radius, due to the form of $P(r)$. The integration over magnitudes is limited by the limiting magnitude $m_{\rm lim}$ of the survey. The normalizations of the radial and magnitude filter produce a background level of unity in the smoothed map (see P96). Therefore, on normalizing, the pixels are expected to have a centrally concentrated distribution about unity, with values exceeding unity representing over-densities.  We obtain pixel distributions covering a large range, with a tail towards positive values. The mode of the distribution is close to unity, and depends on the specific galaxy distribution in the image. \newline

We initially made a catalog of galaxies (objects that have stellarity index less than 0.4) from the optical image, excepting those sources within $200$ image pixels from the edge to avoid these regions of higher image noise.  Smoothed images with grid size between half the core radius to twice the core radius yields similar results (see P96); therefore, we choose to compute the summation above on a grid of pixels spaced by a distance corresponding to the core radius. \newline

\subsection{Parameters quantifying radio source environment}
\label{Subsection_Params}
To examine the environments of the radio sources, we have constructed parameters which quantify the environmental over-density and its distribution in the vicinity of the radio source. For each source, we define a radio axis vector whose direction is taken to be the direction of the longer radio lobe. The angle made by the longer radio lobe with the east-west direction, measured from north to east is designated as the PA (position angle) of the source. For Wide Angle Tailed (WAT) and Head-Tail (HT) sources, the bisecting direction instead of the direction of the larger lobe is used to determine the radio axis in this study. With the radio axis as reference, the smoothed map is resampled.\newline

A circle of $0.5$~Mpc radius is constructed centered at the host galaxy of the radio source, and this circular region is further divided into annular rings $100$~kpc wide.  Along the circumference of each annular region $16$ new equidistant grid-pixels are generated at constant angular distance from each other and at constant distance from the host galaxy. The smoothed image values at the new grid-pixels are calculated by interpolating using neighboring pixels from the original smoothed map. For the annular region defined by each ring $5$ quantities are calculated:
\begin{equation}
a_{k} =  \frac{\sum S_{i}\, f_{k}}{\bar{a_{1}}} ,
\end{equation}
where the summation is over index `i'; {\it i.e.,} over the new grid pixels in the annular region. The functions $f_{k}$ that weight the values of the pixels are $1,\,\mathrm{sin}(\theta_{i}),\,\mathrm{cos}(\theta_{i}),\,\mathrm{sin}(2\,\theta_{i}),\,\mathrm{cos}(2\,\theta_{i})$ for, respectively, $k$ = 1, 2, 3, 4 and 5. The argument of the functions is given by $\theta_{i} = \theta_{ia}+\pi/2$, where the angle $\theta_{ia}$ is the angle of the i\textsuperscript{th} grid-pixel as measured from the radio axis defined above and $S_{i}$ is the value at the grid pixel. The first quantity simply gives a measure of over-density in the environment of the radio source. The other four quantities provide information regarding the dipole and quadrupole anisotropy in the environment of the source. All the quantities are normalized by the average of the $a_{1}$ estimated for the different annuli for each source.\newline

\begin{figure*}
\subfigure[Parameter $a_{2}$]{
\fbox{\includegraphics[scale=0.25]{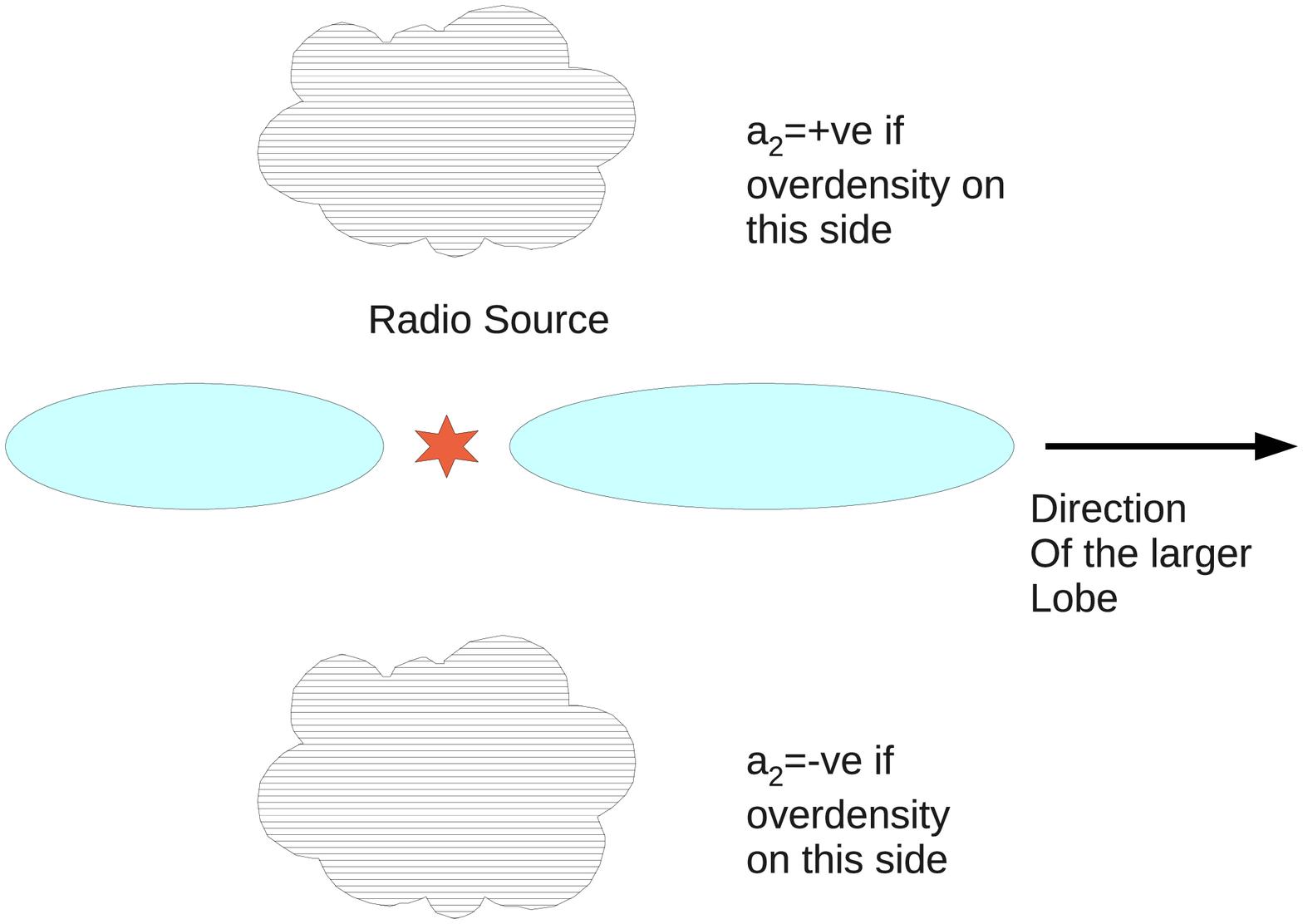}}
\label{lab_a2}
}
\subfigure[Parameter $a_{3}$]{
\fbox{\includegraphics[scale=0.25]{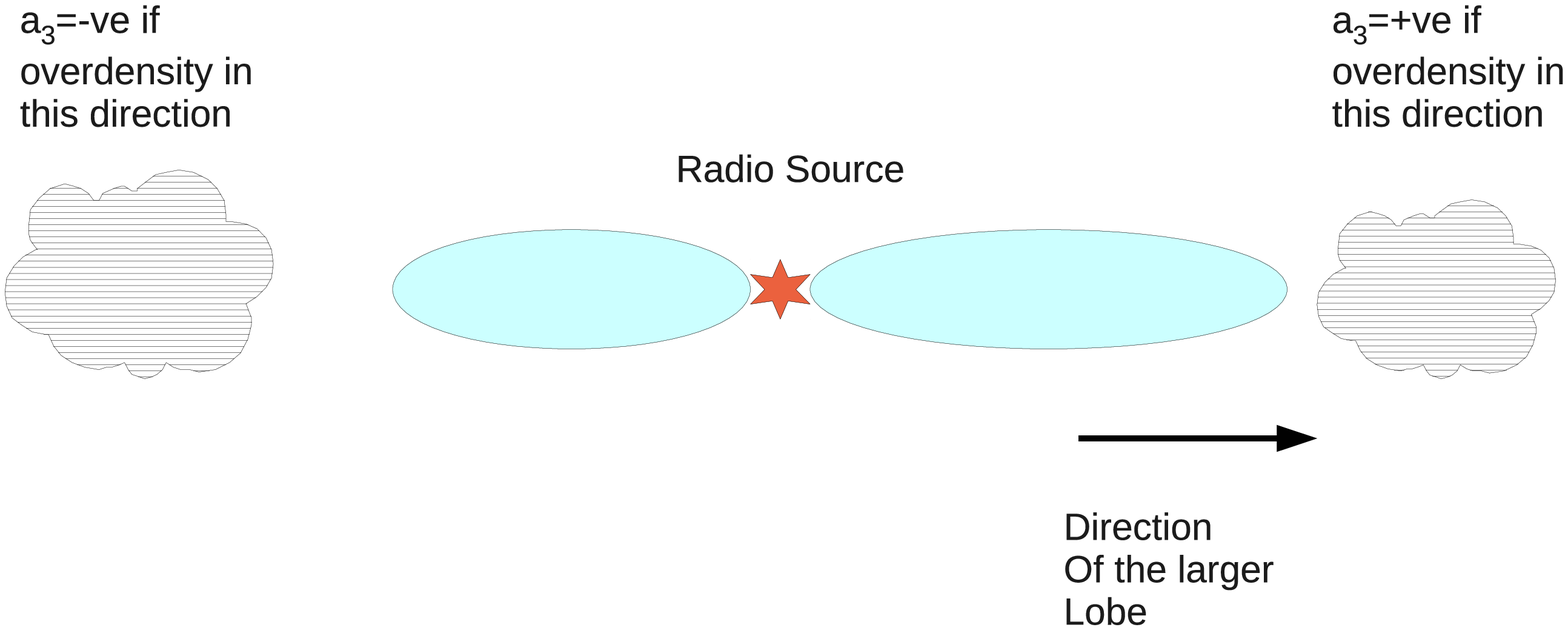}}
\label{lab_a3}
}
\subfigure[Parameter $a_{4}$]{
\fbox{\includegraphics[scale=0.25]{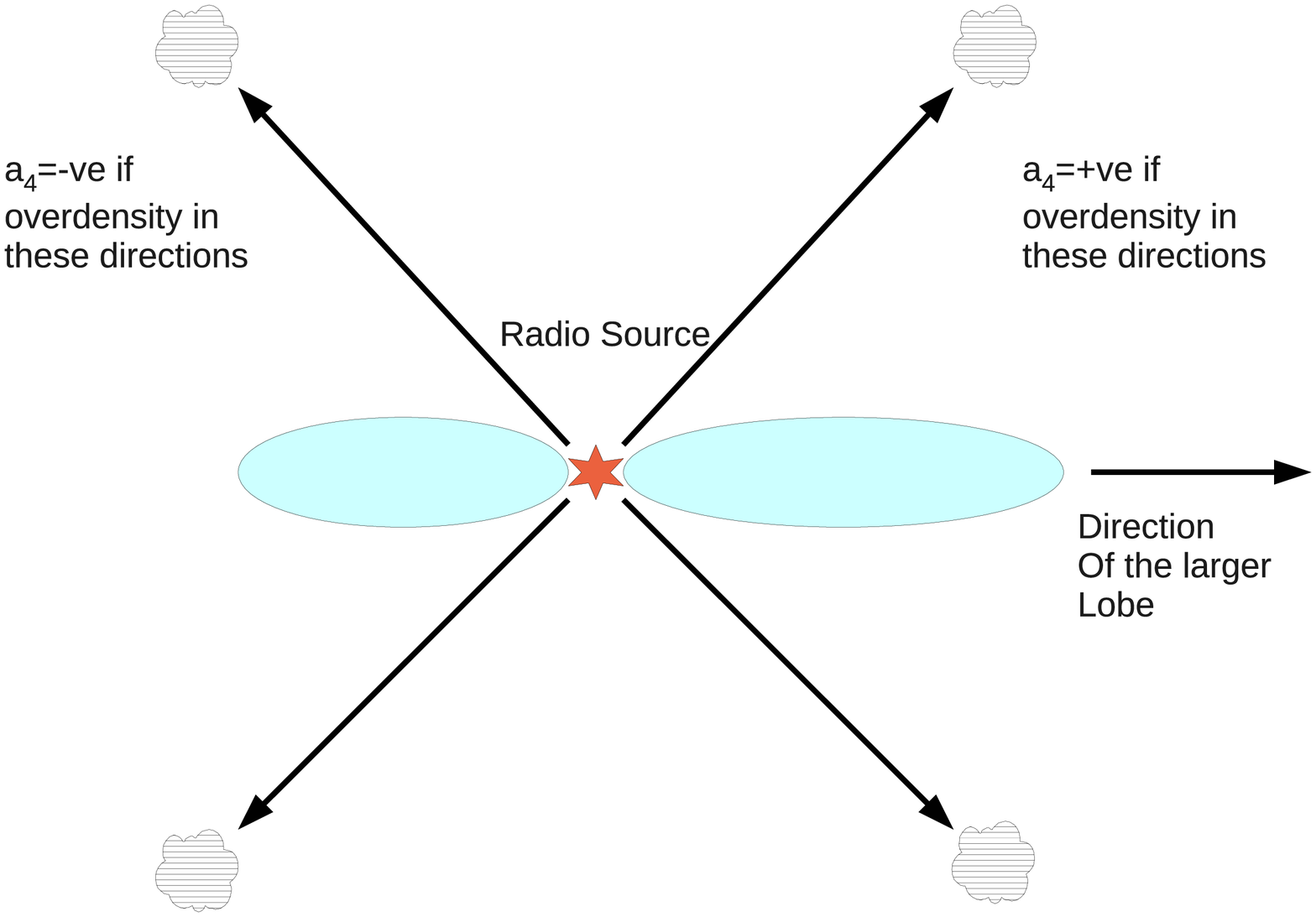}}
\label{lab_a4}
}
\subfigure[Parameter $a_{5}$]{
\fbox{\includegraphics[scale=0.25]{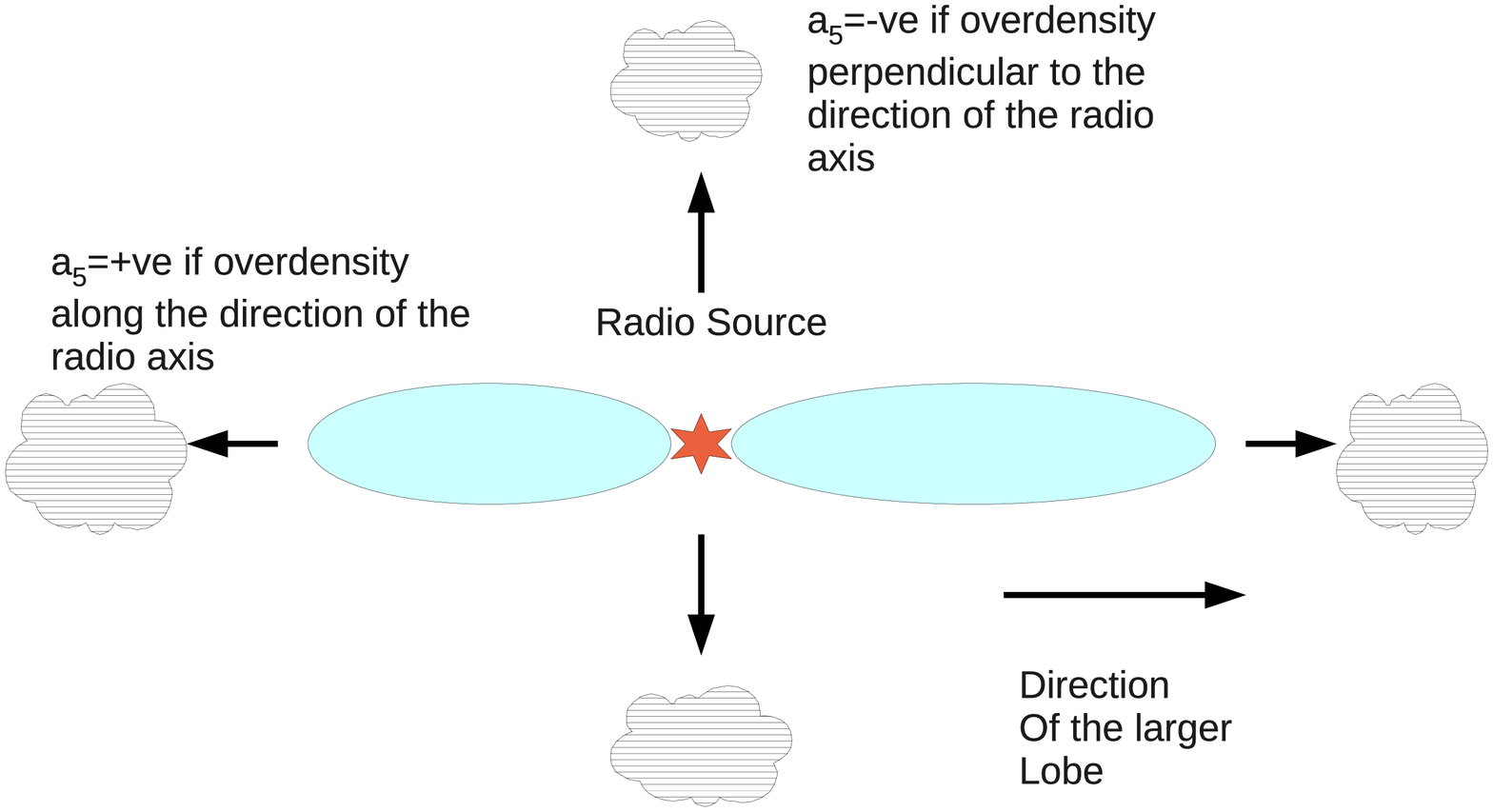}}
\label{lab_a5}
}
\caption{The figure above give schematic depiction of the overdensity parameters $a_{2}$-$a_{5}$ (Panels a-d).}
\label{lab_papar}
\end{figure*}

The `a' parameters represent Fourier components of the angular distribution of galaxy overdensity, or more specifically, the amplitudes of a Fourier harmonic decomposition of angular distributions in galaxies about the radio axis. A schematic depiction of the parameters $a_{2}$ -- $a_{5}$ is given in Fig.~\ref{lab_papar}.\newline

The $a_{1}$ parameter is the mean overdensity and is the amplitude of the zeroth Fourier component. The $a_{2}$ and $a_{3}$ parameters are the fractional side-to-side asymmetry in the galaxy distribution; if the angular distribution follows a dipolar asymmetry then this quantity is unity and the two coefficients $a_{2}$ and $a_{3}$ as well as their signs give the direction of the dipole in the 2D sky plane. The $a_{4}$ and $a_{5}$ parameters are the quadrupolar anisotropy and are the Fourier components of the next order terms.\newline

The errors in these parameters for any source are calculated by sampling different regions of the smoothed map containing the source and constructing the parameters $a_{k}$ in those randomly offset regions. This procedure is repeated at $100$ random positions offset from each source. As shown above, the quantities $a_{k}$ are weighted sums of the pixel values in the vicinity of each source. Therefore, we choose to normalize each quantity by the mean of the first parameter $a_{1}$, which represents the average over-density in the smoothed map. The standard deviations of the five parameters obtained by the above process is also normalized by this $\bar{a_{1}}$. The parameters $a_{k}$ and their standard deviations for all the sample sources are listed in Table~\ref{od_sample}.

\section{Results and Discussion}
\label{Section_Results}
The sources in our sample are divided morphologically into multiple classes. The main classification scheme adopted
is the Fanaroff-Riley classification \citep{FR74}.  Wide-Angle Tailed (WAT) and Head-Tail 
(HT) sources are in separate classes. For a discussion of the classification of the sample sources, see \cite{SST11}. Below we 
describe the results for each of the classes. \newline

The errors in the derived $a_k$ parameters for the individual sources are indeed substantial (see Table~\ref{od_sample}).  Therefore, while estimating the environment for samples of sources of a particular class, we have improved the confidence by computing a weighted mean of each parameter over the sources in the class: this is equivalent to a stacking of images with a weighting corresponding to the noise in the individual images.  We also compute the errors in the weighted means. \newline

\subsection{The environments of Head-Tail and Wide-angle Tailed ATLBS-ESS sources}
\label{Subsection_Env_WATHT}
Wide angle tailed and Head-tail sources are radio galaxies that show extensive signs of `disturbed' radio 
morphology. These sources have bent radio jets/lobes. It has been a long held view that the WAT/HT morphology is a 
result of the interaction of the radio source with cluster gas, either because of ram pressure forces during the movement of the host
galaxy through the cluster gas \citep{OR76} or owing to intra-cluster gas weather created in cluster mergers \citep{BLR02}. The association of these sources with cluster environments has been taken advantage of to detect galaxy clusters at high redshifts 
\citep{BLMR03}. We expect, therefore, that the WAT/HT sources show evidence of inhabiting rich environments.\newline

In our sample, there are $11$ WAT/HT sources. Of the eleven sources, six appear to lie in relatively rich environments 
(showing values above 2.0 for the parameter $a_{1}$, which gives a measure of the `average' overdensity in the 
environment of the source). All the four HT sources in our sample are at relatively low redshifts (below redshift of 0.3)
and all four are found in rich environments with weighted mean of $2.465(\pm 0.197) $ for the $a_{1}$ parameter. The 7 WAT sources 
also have a high overdensity, with a weighted mean value of $1.969(\pm 0.145)$ for $a_{1}$ indicating overdense environments as
expected for this class of sources. There is a hint of decreasing overdensity with redshift 
suggesting that the WATs do not appear to be 
constrained to overdense regions at higher redshifts. However given the difficulties in detecting faint galaxies at higher redshifts 
completeness in galaxy counts will certainly be affected at faint galaxy magnitudes ($m_{r} > 22.75$) and this limits the confidence in the finding of 
any trend with redshift.   
Nevertheless,  the finding that WAT and HT sources do indeed inhabit relatively rich environments is consistent with previous findings and lends confidence in the new method proposed herein for studies of environments of radio sources. A histogram of the values of $a_{1}$ parameter for WAT/HT subsample is presented in Fig.~\ref{lab_watht}. \newline

\subsection{The environments of FR I and FR II ATLBS-ESS sources}
\label{Subsection_Env_FR}
FRI sources have been known to inhabit rich environments: this property has been established with greater confidence for FRIs at relatively lower redshifts. In contrast, FRII sources have been known to favor sparse environments at low redshifts, and are known to reside in richer environments at higher ($z>0.5$) redshifts (Hill and Lilly 1991; Zirbel 1997). Thus FRII radio sources present a remarkable change in their environments with cosmic epoch. Below we discuss the findings from our work for these two classes of sources in our sample. \newline

There are 17 FRI and 15 FRII sources in our sample. The weighted mean value of $a_{1}$ for the 17 FRIs is $1.326 (\pm 0.08)$ where as the corresponding value for the 15 FRIIs is $1.294 \pm 0.098$. The sample includes $4$ FRI sources that have $a_{1}$ below 1. These sources are J0026.8$-$6643, J0026.4$-$6721, J0049.3$-$6703, and J0059.6$-$6712. These sources are not at particularly high redshifts (all of them have redshifts less than 0.5). However, in case of the latter two sources, it is possible that imaging artifacts and the presence of bright stars in the vicinity may have played a role in underestimating the environmental overdensity. \newline

We have separated our subsamples of FRI and FRII sources into two redshift regimes, one below redshift $0.5$ and another above and we compare their environments in each of these regimes. Environments of FRI sources at high redshifts have remained unexplored because of sensitivity issues: FRI sources are lower in luminosity and have more diffuse structures making their detection more difficult  at high redshifts. ATLBS is a survey with high surface brightness sensitivity and has imaged sky regions with good resolution; the ATLBS has detected several FRI sources with $z>0.5$ \citep{SST11}. In our sample of 17 FRIs the redshifts range from 0.21 to 0.97 and there are 4 at redshifts above 0.5. We may, therefore, attempt an examination of the environments of these FRI sources at relatively high redshifts.\newline

However before we examine the results we remind ourselves that because of the finite limit to the sensitivity neighboring galaxy counts at higher redshifts will be progressively underestimated. We accordingly interpret our results, and emphasize that the environments of high redshift sources may only be a lower limit. \newline

We find that at redshifts above $0.5$,  FRI and FRII sources inhabit environments that are not too dissimilar in richness. Moreover, both FRI and FRII type sources are found to lie in relatively overdense environments in the $z>0.5$ regime.  There are $9$ FRII sources and $5$ FRI sources with $z>0.5$ . However, we have not considered the FRI source J0059.6$-$6712 in this comparison as it is on the redshift boundary. Two FRII sources J0105.7$-$6609 and J0057.7$-$6655 for which the results may have been affected by nearby bright stars, as well the FRII source J0105.0$-$6608 for which the host galaxy identification is unclear and the source J0056.6$-$6743 which displays hybrid morphology have been excluded from the exercise. This gives 5 FRII sources and 4 FRI sources for the exercise.  
The average richness (as quantified by the parameter $a_{1}$) for $z>0.5$ FRI sources is $1.335 \pm 0.196$ as compared to an average value of $1.305 \pm 0.177$ for the FRII sources at high redshift. Since we are comparing both groups in the same redshift regime, underestimation of galaxy counts will affect both similarly. \newline

\begin{figure*}
\subfigure[]{
\includegraphics[scale=0.5,angle=270]{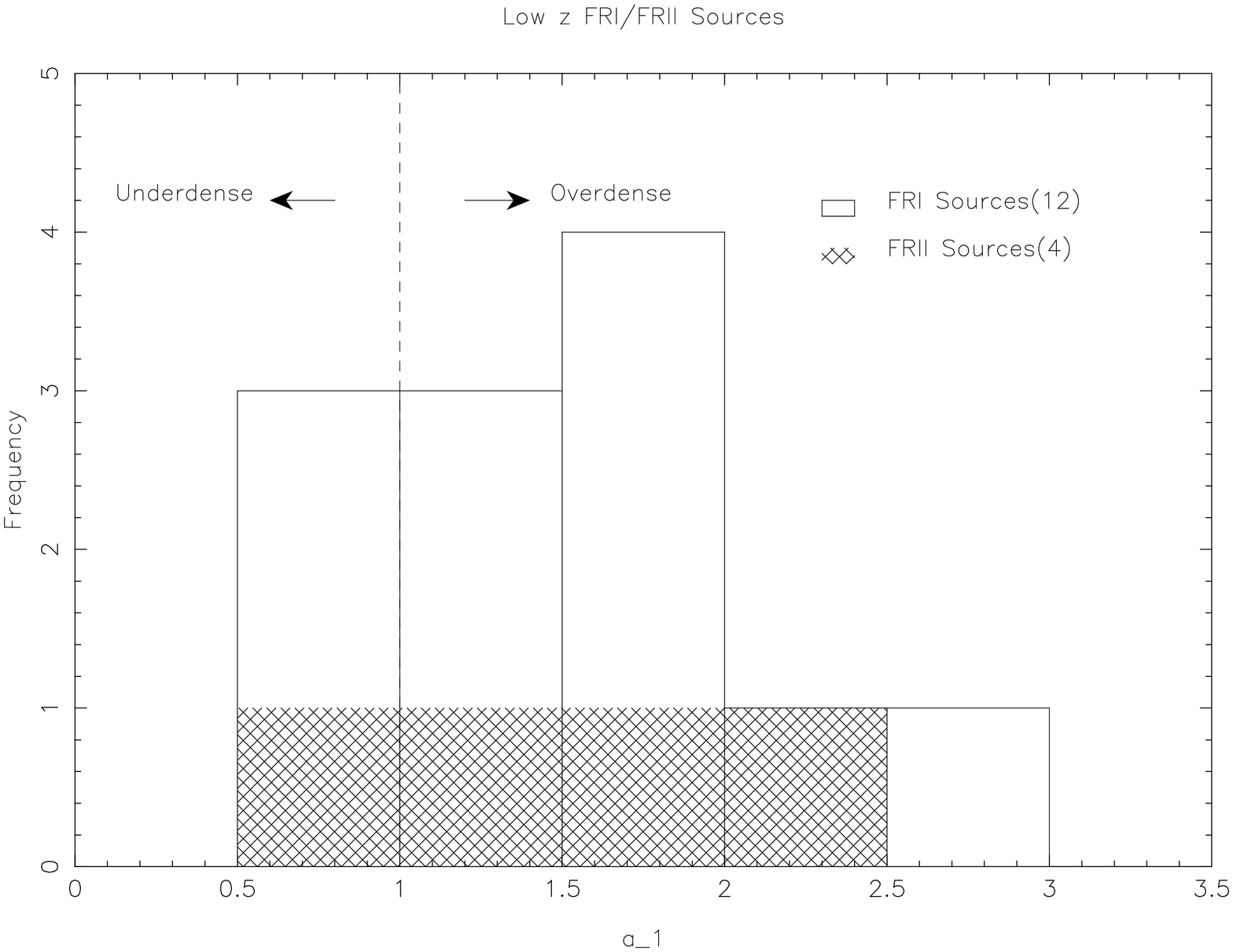}
\label{lab_FRLz}
}
\subfigure[]{
\includegraphics[scale=0.5,angle=270]{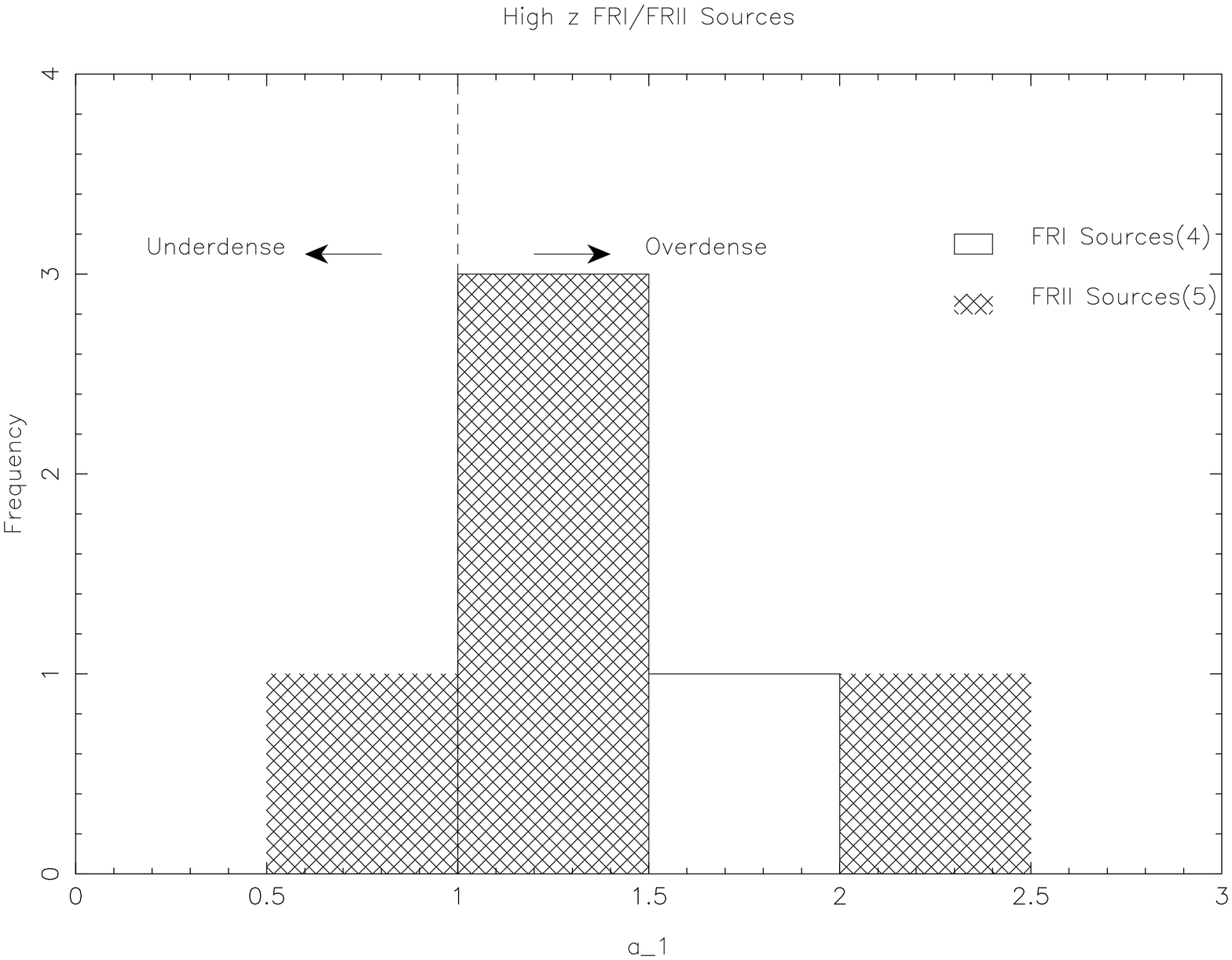}
\label{lab_FRHz} 
}
\caption{Comparative view of the overdensity (as characterized by the parameter $a_{1}$) of (a) low redshift FRI v/s FRII sources and (b) high redshift FRI and FRII sources.}
\label{lab_FRLzHz}
\end{figure*}

FRI sources at low redshifts appear to inhabit a variety of environments: their $a_1$ parameters cover the range 0.68 to 2.57.  
The overall average value of $a_{1}$ for low redshift FRI sources is $1.39 \pm 0.094$ indicating that at $z < 0.5$ FRI sources, 
as expected, generally prefer the relatively higher density environments. Of the $12$ low-redshift FRIs, only $3$ sources are 
in underdense environments and these have an average value of $0.86$ for $a_{1}$.   Our sample includes only a small number of 
FRII sources at low redshifts (only 4 sources in all). We have omitted the source J0046.2$-$6637 for which the identification of 
the host galaxy is uncertain and the source J0044.3$-$6746 that has a bright star nearby, leaving $4$ FRII sources at low redshifts. 
$3$ of the $4$ of these low-$z$ FRIIs are found to have relatively rich environs, with average value of $1.33 \pm 0.159$ for $a_{1}$.  
We note that the environmental richness  parameter $a_1$ is similar for the low redshift FRIs compared to the low redshift FRIIs 
in our ATLBS samples. The above comparison between FRI/FRII sources at high and low redshift is depicted graphically in Fig.~\ref{lab_FRLzHz}.\newline

\begin{figure*}
\subfigure[]{
\includegraphics[scale=0.5,angle=270]{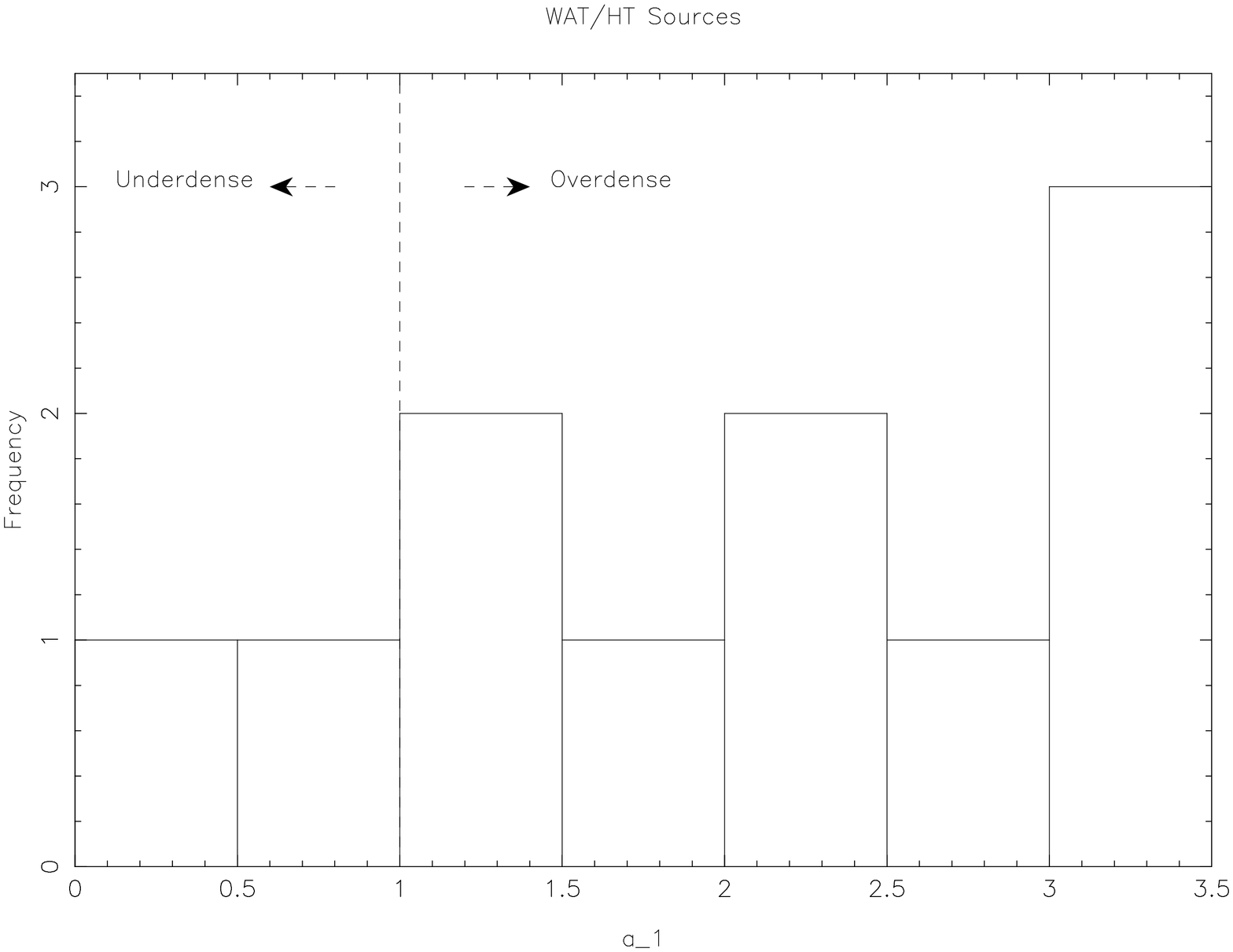}
\label{lab_watht}
}
\subfigure[]{
\includegraphics[scale=0.5,angle=270]{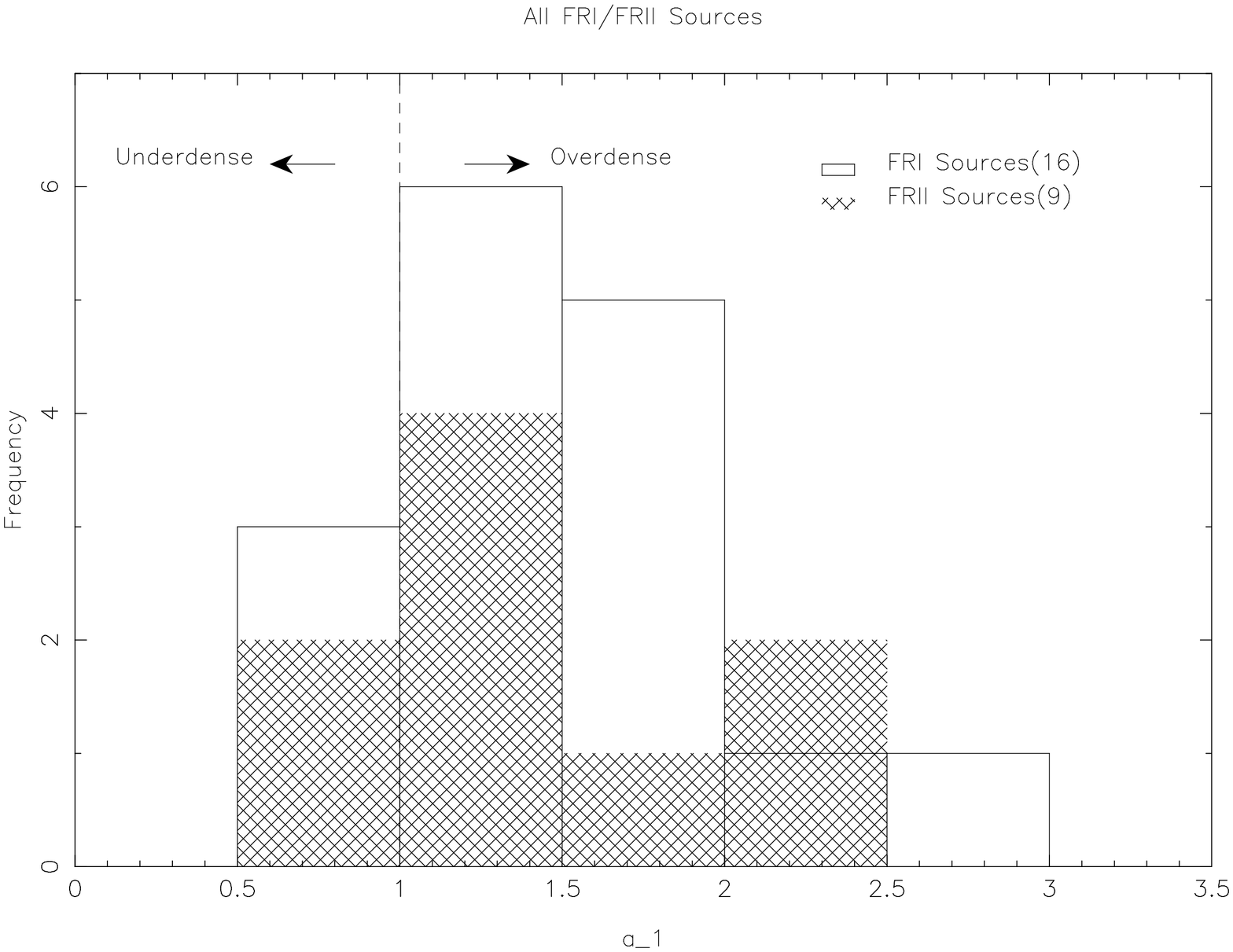}
\label{lab_FRAll}
}
\caption{Histograms of $a_{1}$ parameter for (a) WAT/HT and (b) FRI/FRII sources.}
\label{lab_wathtfrall}
\end{figure*}

An examination of the FRI subsample as a whole reveals that the FRI sources inhabit environments that are more or less similar, 
over the redshift range examined here, barring the extreme outliers.  And a similar result appears to emerge for FRIIs as well: 
their subsamples formed above and below redshift of 0.5 display similar $a_1$ coefficients on the average see Fig.~\ref{lab_FRAll} and Fig.~\ref{lab_FRLzHz}. The weighted mean values of $a_1$ parameter for the selected 16 FRIs and 9 FRIIs (Fig.~\ref{lab_FRAll}) are $1.38 \pm 0.08$ and $1.32 \pm 0.12$. It may be noted here 
that we have separated the HT and WAT sources from this comparison: most HT and WAT sources are FRIs and these clearly lie in more 
overdense regions compared to our FRI sample (which has HT and WAT sources excluded).  Our study suggests that the FRIs and FRIIs 
may have similar environments and occur in moderately overdense galaxy distribution space within galaxy groups and filaments of 
the large scale structure; however, the WAT and HT sources inhabit the more extreme overdensities of clusters of galaxies.  
As is expected in structure formation, the highest density regimes that include clusters of galaxies evolve most rapidly at low 
redshifts and, therefore, it is unsurprising that redshift evolution across a $z=0.5$ boundary appears to be significant only 
for the WAT sources.  \newline

\subsection{Dipole and quadruple environmental anisotropy}
\label{Subsection_Dipole_Quad_Anisotropy}
Next, we examine the environmental parameters $a_{2}$ -- $a_{5}$. These provide information regarding the dipole and quadrupole $\it {angular}$ distribution of the overdensity in the vicinity of the sample sources. If the distribution of the overdensity about the radio source is uniform, then these parameters would be expected to vanish. If the distribution is nonuniform, then the parameters may have non-zero values and the sign of each parameter gives further information regarding the angular distribution. In practice, the value of the parameter is compared with the standard deviation for that parameter to estimate the significance. The arguments of the weighting functions $f_k$ are the angles of the points in the grid with respect to the radio axis.\newline

$a_2$ and $a_3$ measure the dipole anisotropy in density distribution.  The parameter $a_{2}$, which is the overdensity weighted with a sine function, is a measure of the side-to-side density difference on the two sides of the radio axis.   The sign of $a_2$ indicates which side of the source is overdense, and there is no reason to expect any preference for the sign. The parameter $a_2$ would be expected to average to zero for any population of sources because the sign of this parameter would be equally likely to be positive and negative, although individual sources may have a significant magnitude.  The parameter $a_{3}$ is the integral of the azimuthal variation in overdensity weighted with a cosine function. This parameter is of importance when examining asymmetric sources because it is a measure of the overdensity along the radio axis. A positive sign implies that the density in the direction of the longer radio lobe is higher than that towards the shorter radio lobe, and a negative sign implies the opposite. \newline

$a_4$ and $a_5$ measure the quadrupole anisotropy in galaxy density distribution about the radio source.  The parameter $a_{4}$ is the integral overdensity weighted with a sine function for which the argument is twice the position angle with respect to the source axis. This parameter is a matched filter for a quadrupole angular anisotropy in overdensity  that has maxima or minima at angles of $\pi/4$ and $5\pi/4$ to the radio axis.  A positive sign for this parameter implies that the quadrupole anisotropy has overdensities at angles of $\pi/4$ and $5\pi/4$ from the direction defined by the vector towards the more extended lobe, and a negative sign implies that the overdensity is along $3\pi/4$ and $7\pi/4$.  The last parameter $a_{5}$ is weighted by a cosine function that once again has argument twice the position angle with respect to the source axis. This parameter is sensitive to quadrupole anisotropy in density that has maxima along the radio axis or along a direction perpendicular to the source axis. A positive sign for this parameter implies that the overdensity along the radio axis is larger than off the axis, and a negative sign implies that 
the overdensity in a direction perpendicular to the radio axis is larger. Together, these parameters $a_{2}$--$a_{5}$ provide a good description of the dipole and quadrupole distribution of the density in the environments of radio sources.\newline

The parameter $a_{3}$, which is a measure of the environmental dipolar overdensity along the radio axis, has a value consistent with zero (within errors) for most of the source types except HT sources.  The weighted mean $a_{3}$ is $0.123 \pm 0.046$ (for HT sources), $0.018 \pm 0.02$ (for FRI source sub-sample), $-0.021 \pm 0.027$ (for FRII sources) and $-0.044 \pm 0.037$ (for WAT sources). The weighted mean value of $a_{3}$ is significant only for the HT population and it is notable that the value of $a_{3}$ for all of the HT sources is positive. These suggest that the tail of Head-tail sources preferentially---and in all cases in our ATLBS-ESS subsample of HT sources---points in the direction of higher local galaxy density. This result may be interpreted in several ways. One explanation may be that the host galaxies of these HT sources are orbiting around the cluster centre and currently in projection the hosts are moving away from the cluster centres with the tails pointing back towards the cluster centre. Alternatively, the sources may be in clusters undergoing merger events and the tails of the HT sources are being dragged by the intracluster weather toward the cluster centre. \newline

\subsection{Asymmetric ATLBS-ESS sources}
\label{Subsection_Asymmetric_Sources}
We next examine the sources which exhibit significantly asymmetric radio morphology. A subsample of `asymmetric' sources was compiled on the basis of lobe asymmetry; for inclusion in this sub-sample, one of the lobes is required to be more than 1.5 times the extent of the opposite lobe. With this selection criterion, we find $7$ asymmetric sources in ATLBS-ESS. We have presented the asymmetric source sample separately in Table.~\ref{as_sample}.\newline

We have examined the environmental parameters for the selected asymmetric sources. Almost all the asymmetric sources in our sample appear to lie in rich environments, as indicated by high to moderate values of $a_{1}$; the one exception is  J0101.1$-$6600, which has an $a_{1}$ parameter corresponding to an underdense region. We note that all the asymmetric sources, except J0045.5$-$6726, show negative values for the parameter $a_{3}$, which is a clear indication that the ambient galaxy density is almost always higher in the direction of the shorter lobe.  The weighted mean value for $a_3$ is $-0.0359 \pm 0.0341$. The latter value is significant when taken in the context of the values of $a_{3}$ for the subsample of \textit{symmetric} sources. The values of $a_{3}$ parameter are positive or negative for symmetric sources, without any significant preference towards positive or negative sign. In comparison, all except one of the sources in the asymmetric source sample have a negative sign for the value of the $a_{3}$ parameter. Examined in this light, the distribution of the $a_{3}$ parameter is significant. \newline

All of the asymmetric sources have a positive $a_5$ and the weighted mean $a_5$ for the asymmetric sources is significantly $0.051 \pm 0.013$. This implies that asymmetric radio sources are usually aligned along the line of excess galaxy density with a quadrupole asymmetry apart from any dipole component.  It may be that when double radio sources are created by jets that happen to be aligned with galaxy overdensity, and the galaxy clustering is on one side, the associated gas inhibits jet advance via ram pressure interaction and it is radio galaxies in such environmental circumstances that display grossly asymmetric morphology.  It may also be noted that none of the subsamples of sources, except the asymmetric sources show significant values of $a_{5}$.  In most of the asymmetric cases, there is more galaxy overdensity on the shorter side, consistent with the expectations that gas density follows galaxy density and the side with higher gas density would be 
expected to be shorter owing to slower advance speed for the jets. The positive value of $a_{5}$, together with a negative value of $a_{3}$ for most of the sample implies that the environmental overdensity is not a gradient but is a concentration in the direction of the shorter lobe. The $a_4$ parameter for these asymmetric sources has less significance in magnitude and appears random in signs, as expected. Curiously, the parameter $a_2$ for this sample is positive for all of the asymmetric sources; however, the weighted mean $a_2$ is $0.048 \pm 0.034$ and is not significant.   \newline

All of the six asymmetric sources with negative values for $a_{3}$ have linear size in the range 400 -- 600~kpc. J0045.5$-$6726, which alone has a positive value for $a_{3}$, has a significantly smaller linear size of $164$ kpc and, therefore, the anomalous behavior for this source may be understood as arising from its relatively small size due to which the lobe extent may be more influenced by the local inter-stellar medium of the host galaxy rather than the intergalactic gas associated with the large scale galaxy distribution, which is what is probed by our approach.  The high value of $a_{1}$ for the FRI source J0045.5$-$6726 indicates a rich environment for this source; however, the positive sign for $a_{3}$ implies that the longer side of the source is in denser regions. Additionally, the source has a high value for $a_{5}$, showing that the density distribution is along the radio axis.  This source has an FRI morphology and an alternate explanation for the positive value of $a_3$ for this source may be that the higher galaxy density on the longer side preserves the diffuse emission on that side by limiting expansion losses. \newline

\section{Summary}
\label{Summary}
We have presented details of optical observations and data reduction for the ATLBS survey regions. The optical observations were used 
to determine the redshifts and magnitudes of the ATLBS-ESS sample of radio sources formed of the extended radio sources detected in 
the ATLBS survey. In this study, galaxy catalogs constructed from the optical data were utilized to estimate the environmental 
parameters of selected sources from the ATLBS-ESS sample.  We have defined a set of parameters $a_{1}$ -- $a_{5}$ to quantify the local 
galaxy overdensity and its angular anisotropy with respect to the axis of the radio sources using smoothed galaxy density maps.  
Dipole and quadrupole anisotropy has been estimated for the individual sources and these measures have been stacked (averaged) to 
estimate mean measures and their errors for different classes of radio source morphologies. \newline  

Examining the anisotropy parameters for a sub-sample of extended double radio sources that includes all sources with pronounced 
asymmetry in lobe extents, we find good evidence for environmental anisotropy being the dominant cause for lobe asymmetry in that 
higher galaxy density occurs almost always on the side of the shorter lobe, and this validates the usefulness of the method proposed 
and adopted here.  The environmental parameters have been used to examine and compare the environments of FRI and FRII 
radio sources in two redshift regimes ($z<0.5$ and $z>0.5$).  Wide-angle tail sources and Head-tail sources lie in the most overdense 
environments.  The Head tail source environments display dipolar anisotropy in that higher galaxy density appears to lie in 
the direction of the tails.  Excluding the Head-tail and Wide-angle tail sources, subsamples of FRI and FRII sources 
from the ATLBS survey appear to lie in similar moderately overdense environments, with no evidence for redshift evolution in the 
regimes studied herein.

\section*{Acknowledgements}
\label{Section_Acknowledgement}
KT and LS would like to thank Dr. Helen Johnston for providing the spectroscopic data. \newline

IRAF is distributed by the National Optical Astronomy Observatories, which are operated by the Association of Universities for Research in Astronomy, Inc., under cooperative agreement with the National Science Foundation.

\newpage

\begin{landscape}
\begin{table}
\caption{\scriptsize The table presents the sample along with the parameters $a_{k}$ and the standard deviations in each of the parameters. Column 1 gives the source name, with the extended designation given in column 2 (where ** denotes a source designation taken from the Sydney University Molonglo Sky Survey (SUMSS) \citep{Mau03}) Column 3 and 4 give the redshift of the source and the source type respectively, with '*' denoting spectroscopically measured values for the redshift. Column 5 gives the position angle of the sources in degrees. Columns 6-14 present each overdensity parameter and the normalized standard deviations for each parameter. The column for each of the overdensity parameter gives three values: the best estimate of the value of the parameter itself as well as the values it may have at $\pm 5$ degrees from the listed PA, giving the uncertainty in the parameter due to the uncertainty in PA (The angle made by the longer radio lobe 
with the east-west direction, measured from north to east is designated as the position angle
of the source. For WAT-HT sources, the bisecting direction instead of the direction of the 
longer lobe is used to determine the radio axis in this study).}
\label{od_sample}
\begin{tabular}{@{}llclcccccccccccc@{}}
\hline

Source Name & Extended Source Name & Redshift & Source Type & PA & $a_{1}$ & $\sigma_{a_{1}}$& $a_{2}$& $\sigma_{a_{2}}$& $a_{3}$& $\sigma_{a_{3}}$ & $a_{4}$ & $\sigma_{a_{4}}$ & $a_{5}$ & $\sigma_{a_{5}}$ \\
\hline
J0024.4$-$6636	& J002426$-$663612 &	0.21	&	I	&	90	&	$	2.57^{2.57}_{2.57}$	&	0.4	&	$	0.09^{0.1}_{0.07}$&	0.1	&	$	-0.15^{-0.14}_{-0.16}$&	0.11	&	$	0.03^{0.02}_{0.04}$&	0.04	&	$	0.04^{0.04}_{0.03}$&	0.04	\\
J0025.0$-$6658	& J002500$-$665804**&	0.64	&	WAT	&	27	&	$	1.31^{1.31}_{1.31}$	&	0.37	&	$	-0.14^{-0.14}_{-0.14}$&	0.1	&	$	-0.01^{-0.02}_{0}$&	0.11	&	$	-0.01^{-0.01}_{0.05}$&	0.05	&	$	-0.01^{-0.01}_{0}$&	0.04	\\
J0026.4$-$6721	& J002628$-$672148&	0.27*	&	I	&	-90	&	$	0.97^{0.97}_{0.97}$	&	0.25	&	$	0.09^{0.08}_{0.09}$&	0.07	&	$	0.08^{0.09}_{0.07}$&	0.07	&	$	0.02^{0.02}_{0.03}$&	0.03	&	$	0.03^{0.03}_{0.02}$&	0.03	\\
J0026.8$-$6643	& J002648$-$664402** &	0.21	&	I	&	207	&	$	0.93^{0.93}_{0.93}$	&	0.4	&	$	-0.07^{-0.07}_{-0.07}$&	0.1	&	$	-0.01^{-0.01}_{0}$&	0.1	&	$	-0.05^{-0.05}_{0.04}$&	0.04	&	$	0^{-0.01}_{0.01}$&	0.04	\\
J0030.7$-$6714	&J003045$-$671437**  &	0.97	&	I	&	37	&	$	1.07^{1.07}_{1.07}$	&	0.59	&	$	0.13^{0.12}_{0.13}$&	0.21	&	$	0.08^{0.09}_{0.07}$&	0.2	&	$	-0.06^{-0.06}_{0.11}$&	0.11	&	$	0.01^{0}_{0.02}$&	0.09	\\
J0031.1$-$6642	& J003108$-$664245**  &	0.77	&	II	&	-6	&	$	0.65^{0.65}_{0.65}$	&	0.47	&	$	-0.07^{-0.07}_{-0.06}$&	0.13	&	$	0.06^{0.06}_{0.07}$&	0.14	&	$	0.02^{0.03}_{0.05}$&	0.06	&	$	-0.03^{-0.03}_{-0.03}$&	0.06	\\
J0032.9$-$6614	&  J003257$-$661417** &	0.92*	&	I	&	-34	&	$	1.63^{1.63}_{1.63}$	&	0.58	&	$	-0.55^{-0.57}_{-0.53}$&	0.19	&	$	0.2^{0.15}_{0.25}$&	0.19	&	$	-0.06^{-0.05}_{0.09}$&	0.09	&	$	-0.04^{-0.05}_{-0.03}$&	0.09	\\
J0033.4$-$6714	&  J003329$-$671415** &	0.41*	&	I	&	239	&	$	1.33^{1.33}_{1.33}$	&	0.25	&	$	0.1^{0.09}_{0.11}$&	0.06	&	$	0.14^{0.15}_{0.13}$&	0.06	&	$	0.1^{0.11}_{0.03}$&	0.03	&	$	-0.06^{-0.04}_{-0.08}$&	0.03	\\
J0035.0$-$6612	& J003501$-$661252** &	0.47*	&	I	&	187	&	$	1.6^{1.6}_{1.6}$	&	0.32	&	$	0.12^{0.12}_{0.12}$&	0.07	&	$	0^{0.01}_{-0.01}$&	0.07	&	$	-0.05^{-0.05}_{0.04}$&	0.04	&	$	0^{-0.01}_{0.01}$&	0.03	\\
J0036.9$-$6645	& J003654$-$664513  &	0.23	&	I	&	63	&	$	1.21^{1.21}_{1.21}$	&	0.34	&	$	-0.07^{-0.06}_{-0.08}$&	0.1	&	$	-0.08^{-0.09}_{-0.08}$&	0.1	&	$	0.03^{0.03}_{0.04}$&	0.04	&	$	-0.04^{-0.04}_{-0.05}$&	0.04	\\
J0040.2$-$6553	& J004014$-$655325  &	0.51*	&	II	&	243	&	$	1.3^{1.3}_{1.3}$	&	0.3	&	$	-0.07^{-0.08}_{-0.07}$&	0.09	&	$	0.05^{0.04}_{0.06}$&	0.09	&	$	0.02^{0.03}_{0.03}$&	0.03	&	$	-0.02^{-0.02}_{-0.03}$&	0.04	\\
J0041.7$-$6726	& J004145$-$672629** &	0.29*	&	HT	&	256	&	$	3.32^{3.32}_{3.32}$	&	0.45	&	$	-0.02^{-0.03}_{-0.02}$&	0.09	&	$	0.06^{0.06}_{0.06}$&	0.09	&	$	0.02^{0.02}_{0.03}$&	0.03	&	$	-0.02^{-0.01}_{-0.02}$&	0.03	\\
J0042.1$-$6728	& J004208$-$672805**  &	0.25	&	HT	&	180	&	$	2.97^{2.97}_{2.97}$	&	0.48	&	$	0.23^{0.22}_{0.25}$&	0.09	&	$	0.18^{0.2}_{0.16}$&	0.11	&	$	-0.04^{-0.02}_{0.03}$&	0.03	&	$	-0.14^{-0.15}_{-0.13}$&	0.03	\\
J0043.2$-$6751	& J004317$-$675147**  &	0.38	&	WAT	&	90	&	$	3.01^{3.01}_{3.01}$	&	0.32	&	$	-0.06^{-0.08}_{-0.05}$&	0.09	&	$	0.19^{0.18}_{0.19}$&	0.08	&	$	0.07^{0.07}_{0.04}$&	0.04	&	$	0^{0.01}_{-0.02}$&	0.04	\\
J0043.6$-$6624	& J004337$-$662447 &	0.99	&	WAT	&	124	&	$	0.46^{0.46}_{0.46}$	&	0.74	&	$	0.05^{0.05}_{0.05}$&	0.25	&	$	0^{0.01}_{0}$&	0.27	&	$	0.01^{0}_{0.13}$&	0.13	&	$	0.05^{0.05}_{0.05}$&	0.11	\\
J0044.3$-$6746	& J004419$-$674657**  &	0.29	&	II	&	198	&	$	1.73^{1.73}_{1.73}$	&	0.36	&	$	0.19^{0.19}_{0.2}$&	0.1	&	$	0.04^{0.06}_{0.02}$&	0.1	&	$	0.02^{0.03}_{0.04}$&	0.04	&	$	-0.06^{-0.05}_{-0.06}$&	0.04	\\
J0044.7$-$6656	& J004451$-$665628**  &	0.72	&	II	&	55	&	$	2.02^{2.02}_{2.02}$	&	0.4	&	$	-0.19^{-0.19}_{-0.19}$&	0.11	&	$	0.05^{0.03}_{0.06}$&	0.11	&	$	-0.03^{-0.03}_{0.05}$&	0.05	&	$	-0.02^{-0.03}_{-0.02}$&	0.05	\\
J0045.5$-$6726	& J004532$-$672635** &	0.27	&	I	&	72	&	$	1.93^{1.93}_{1.93}$	&	0.46	&	$	0.01^{-0.01}_{0.03}$&	0.1	&	$	0.21^{0.21}_{0.21}$&	0.09	&	$	0.01^{-0.01}_{0.03}$&	0.03	&	$	0.13^{0.13}_{0.13}$&	0.03	\\
J0046.2$-$6637	& J004613$-$663708** &	0.37	&	II	&	135	&	$	1.75^{1.75}_{1.75}$	&	0.36	&	$	0.2^{0.19}_{0.21}$&	0.08	&	$	0.09^{0.11}_{0.07}$&	0.11	&	$	0.05^{0.05}_{0.05}$&	0.05	&	$	-0.01^{0}_{-0.02}$&	0.04	\\
J0049.3$-$6703	& J004922$-$670358 &	0.47	&	I	&	240	&	$	0.68^{0.68}_{0.68}$	&	0.29	&	$	-0.04^{-0.04}_{-0.04}$&	0.08	&	$	0^{0}_{0.01}$&	0.07	&	$	-0.01^{0}_{0.03}$&	0.03	&	$	-0.03^{-0.03}_{-0.03}$&	0.03	\\
J0052.7$-$6651	& J005248$-$665109**   &	0.24	&	HT	&	-45	&	$	2.39^{2.39}_{2.39}$	&	0.44	&	$	-0.04^{-0.06}_{-0.03}$&	0.1	&	$	0.19^{0.19}_{0.19}$&	0.1	&	$	0.06^{0.06}_{0.04}$&	0.05	&	$	-0.01^{0}_{-0.03}$&	0.04	\\
J0055.7$-$6610	& J005548$-$661031**  &	0.22	&	HT	&	56	&	$	1.92^{1.92}_{1.92}$	&	0.3	&	$	-0.06^{-0.07}_{-0.05}$&	0.1	&	$	0.1^{0.1}_{0.11}$&	0.08	&	$	0.01^{0}_{0.03}$&	0.03	&	$	0.07^{0.07}_{0.07}$&	0.04	\\
J0056.4$-$6651	& J005627$-$665122**  &	0.19	&	WAT	&	-45	&	$	2.13^{2.13}_{2.13}$	&	0.44	&	$	0.01^{0.02}_{0}$&	0.09	&	$	-0.11^{-0.11}_{-0.12}$&	0.09	&	$	-0.05^{-0.05}_{0.04}$&	0.04	&	$	0.02^{0.01}_{0.03}$&	0.04	\\
J0056.6$-$6743	& J005637$-$674343**  &	0.93	&	II	&	79	&	$	1.26^{1.26}_{1.26}$	&	0.51	&	$	0.39^{0.38}_{0.4}$&	0.19	&	$	0.15^{0.19}_{0.12}$&	0.2	&	$	0.02^{0.03}_{0.09}$&	0.09	&	$	-0.07^{-0.07}_{-0.07}$&	0.09	\\
J0056.9$-$6632	& J005657$-$663239  &	0.25*	&	II	&	53	&	$	2.03^{2.03}_{2.03}$	&	0.41	&	$	-0.06^{-0.06}_{-0.07}$&	0.07	&	$	-0.04^{-0.05}_{-0.03}$&	0.08	&	$	0.02^{0.01}_{0.03}$&	0.03	&	$	0.05^{0.05}_{0.04}$&	0.03	\\
J0057.0$-$6734	& J005704$-$673413   &	0.31*	&	II	&	-90	&	$	1.73^{1.73}_{1.73}$	&	0.32	&	$	-0.15^{-0.14}_{-0.17}$&	0.07	&	$	-0.15^{-0.16}_{-0.13}$&	0.08	&	$	-0.01^{-0.01}_{0.03}$&	0.03	&	$	-0.01^{-0.02}_{-0.01}$&	0.04	\\
J0057.2$-$6651	& J005707$-$665059 &	0.24*	&	I	&	112	&	$	1.82^{1.82}_{1.82}$	&	0.41	&	$	0.1^{0.1}_{0.09}$&	0.08	&	$	-0.03^{-0.02}_{-0.03}$&	0.08	&	$	-0.03^{-0.04}_{0.03}$&	0.03	&	$	0.04^{0.04}_{0.05}$&	0.03	\\
J0057.4$-$6703	& J005728$-$670325**  &	0.26*	&	WAT	&	225	&	$	3.13^{3.13}_{3.13}$	&	0.35	&	$	-0.05^{-0.04}_{-0.06}$&	0.1	&	$	-0.13^{-0.14}_{-0.13}$&	0.08	&	$	0.06^{0.06}_{0.04}$&	0.04	&	$	0.02^{0.03}_{0.01}$&	0.04	\\
J0057.7$-$6655	& J005745$-$665507 &	0.66	&	II	&	120	&	$	0.79^{0.79}_{0.79}$	&	0.33	&	$	0.03^{0.04}_{0.02}$&	0.09	&	$	-0.07^{-0.07}_{-0.08}$&	0.09	&	$	0.05^{0.04}_{0.05}$&	0.05	&	$	0.07^{0.07}_{0.06}$&	0.05	\\
J0059.6$-$6712	& J005941$-$671254** &	0.5	&	I	&	180	&	$	0.81^{0.81}_{0.81}$	&	0.26	&	$	-0.1^{-0.09}_{-0.09}$&	0.08	&	$	0^{-0.01}_{0.01}$&	0.07	&	$	0^{0}_{0.03}$&	0.03	&	$	0.02^{0.02}_{0.02}$&	0.03	\\
J0101.1$-$6600	& J010107$-$660018 &	0.24	&	II	&	174	&	$	0.81^{0.81}_{0.81}$	&	0.3	&	$	0.02^{0.02}_{0.02}$&	0.07	&	$	-0.01^{-0.01}_{-0.02}$&	0.08	&	$	0.03^{0.02}_{0.03}$&	0.03	&	$	0.04^{0.04}_{0.03}$&	0.03	\\
J0101.5$-$6742	& J010134$-$674214** &	0.59	&	WAT	&	-63	&	$	1.29^{1.29}_{1.29}$	&	0.29	&	$	-0.1^{-0.09}_{-0.12}$&	0.07	&	$	-0.19^{-0.2}_{-0.18}$&	0.08	&	$	-0.02^{-0.02}_{0.03}$&	0.03	&	$	-0.01^{-0.01}_{0}$&	0.04	\\
J0102.6$-$6658	& J010238$-$665813**  &	0.61	&	I	&	40	&	$	1.46^{1.46}_{1.46}$	&	0.4	&	$	0.07^{0.09}_{0.05}$&	0.11	&	$	-0.24^{-0.23}_{-0.25}$&	0.11	&	$	-0.02^{-0.03}_{0.05}$&	0.05	&	$	0.04^{0.03}_{0.04}$&	0.04	\\
J0102.9$-$6722	& J010256$-$672220 &	0.84	&	WAT	&	135	&	$	0.56^{0.56}_{0.56}$	&	0.54	&	$	0.23^{0.22}_{0.23}$&	0.15	&	$	0.06^{0.08}_{0.04}$&	0.16	&	$	0.04^{0.04}_{0.07}$&	0.07	&	$	-0.04^{-0.03}_{-0.04}$&	0.08	\\
J0103.1$-$6632	& J010310$-$663221  &	0.4*	&	I	&	90	&	$	1.2^{1.2}_{1.2}$	&	0.31	&	$	0.02^{0.02}_{0.02}$&	0.07	&	$	-0.01^{-0.01}_{-0.01}$&	0.07	&	$	-0.03^{-0.03}_{0.03}$&	0.03	&	$	0.02^{0.02}_{0.03}$&	0.03	\\
J0103.2$-$6614	& J010315$-$661425 &	0.33*	&	II	&	186	&	$	1.17^{1.17}_{1.17}$	&	0.28	&	$	0.02^{0.02}_{0.02}$&	0.08	&	$	-0.04^{-0.04}_{-0.05}$&	0.08	&	$	0^{0}_{0.03}$&	0.03	&	$	0.01^{0.01}_{0.01}$&	0.03	\\
J0103.7$-$6632	& J010344$-$663227  &	0.59	&	II	&	-90	&	$	1.2^{1.2}_{1.2}$	&	0.37	&	$	0.04^{0.05}_{0.03}$&	0.11	&	$	-0.12^{-0.11}_{-0.12}$&	0.1	&	$	0.02^{0.01}_{0.04}$&	0.04	&	$	0.04^{0.05}_{0.04}$&	0.05	\\
J0103.7$-$6747	& J010345$-$674746**   &	0.33*	&	I	&	-63	&	$	2.26^{2.26}_{2.26}$	&	0.44	&	$	-0.08^{-0.08}_{-0.09}$&	0.08	&	$	-0.05^{-0.06}_{-0.04}$&	0.08	&	$	0.06^{0.06}_{0.03}$&	0.03	&	$	-0.01^{0}_{-0.02}$&	0.03	\\
J0105.0$-$6608	& J010500$-$660856**   &	0.85	&	II	&	248	&	$	0.32^{0.32}_{0.32}$	&	0.59	&	$	-0.04^{-0.03}_{-0.05}$&	0.22	&	$	-0.09^{-0.09}_{-0.09}$&	0.19	&	$	0.02^{0.01}_{0.08}$&	0.08	&	$	0.02^{0.02}_{0.02}$&	0.09	\\
J0105.7$-$6609	& J010540$-$660940** &	0.98	&	II	&	-85	&	$	0.14^{0.14}_{0.14}$	&	0.94	&	$	-0.02^{-0.01}_{-0.02}$&	0.28	&	$	-0.06^{-0.06}_{-0.05}$&	0.31	&	$	0.02^{0.01}_{0.12}$&	0.12	&	$	0.01^{0.02}_{0.01}$&	0.13	\\
J0106.0$-$6653	&  J010601$-$665337  &	0.26*	&	I	&	63	&	$	1.7^{1.7}_{1.7}$	&	0.29	&	$	0.15^{0.15}_{0.15}$&	0.1	&	$	0.02^{0.04}_{0.01}$&	0.09	&	$	0.04^{0.06}_{0.03}$&	0.03	&	$	-0.1^{-0.09}_{-0.1}$&	0.04	\\
J0108.6$-$6655	& J010838$-$665527**  &	0.53*	&	I	&	27	&	$	1.27^{1.27}_{1.27}$	&	0.27	&	$	0.08^{0.07}_{0.08}$&	0.08	&	$	0.08^{0.09}_{0.08}$&	0.09	&	$	0.01^{0.01}_{0.03}$&	0.04	&	$	0^{0}_{0}$&	0.04	\\
J0110.7$-$6705	& J011046$-$670507**  &	0.8	&	II	&	159	&	$	1.03^{1.03}_{1.03}$	&	0.64	&	$	-0.19^{-0.21}_{-0.17}$&	0.22	&	$	0.21^{0.2}_{0.23}$&	0.24	&	$	-0.12^{-0.13}_{0.09}$&	0.09	&	$	0.08^{0.06}_{0.1}$&	0.1	\\

\hline
\end{tabular}
\end{table}
\end{landscape}

\begin{landscape}
\begin{table}
\caption{The table presents the asymmetric source subsample along with the parameters $a_{k}$ and the standard deviations in each of the parameters. Various columns follow the same scheme as that for Table.~\ref{od_sample}.}
\label{as_sample}
\begin{tabular}{@{}llclcccccccccccc@{}}
\hline
Source Name & Extended Source Name & Redshift & Source Type & PA & $a_{1}$ & $\sigma_{a_{1}}$& $a_{2}$& $\sigma_{a_{2}}$& $a_{3}$& $\sigma_{a_{3}}$ & $a_{4}$ & $\sigma_{a_{4}}$ & $a_{5}$ & $\sigma_{a_{5}}$ \\
\hline
J0024.4$-$6636	& J002426$-$663612 &	0.21	&	I	&	90	&	$	2.57^{2.57}_{2.57}$	&	0.4	&	$	0.09^{0.1}_{0.07}$&	0.1	&	$	-0.15^{-0.14}_{-0.16}$&	0.11	&	$	0.03^{0.02}_{0.04}$&	0.04	&	$	0.04^{0.04}_{0.03}$&	0.04	\\
J0045.5$-$6726	& J004532$-$672635** &	0.27	&	I	&	72	&	$	1.93^{1.93}_{1.93}$	&	0.46	&	$	0.01^{-0.01}_{0.03}$&	0.1	&	$	0.21^{0.21}_{0.21}$&	0.09	&	$	0.01^{-0.01}_{0.03}$&	0.03	&	$	0.13^{0.13}_{0.13}$&	0.03	\\
J0057.2$-$6651	& J005707$-$665059 &	0.24*	&	I	&	112	&	$	1.82^{1.82}_{1.82}$	&	0.41	&	$	0.1^{0.1}_{0.09}$&	0.08	&	$	-0.03^{-0.02}_{-0.03}$&	0.08	&	$	-0.03^{-0.04}_{0.03}$&	0.03	&	$	0.04^{0.04}_{0.05}$&	0.03	\\
J0101.1$-$6600	& J010107$-$660018 &	0.24	&	II	&	174	&	$	0.81^{0.81}_{0.81}$	&	0.3	&	$	0.02^{0.02}_{0.02}$&	0.07	&	$	-0.01^{-0.01}_{-0.02}$&	0.08	&	$	0.03^{0.02}_{0.03}$&	0.03	&	$	0.04^{0.04}_{0.03}$&	0.03	\\
J0102.6$-$6658	& J010238$-$665813**  &	0.61	&	I	&	40	&	$	1.46^{1.46}_{1.46}$	&	0.4	&	$	0.07^{0.09}_{0.05}$&	0.11	&	$	-0.24^{-0.23}_{-0.25}$&	0.11	&	$	-0.02^{-0.03}_{0.05}$&	0.05	&	$	0.04^{0.03}_{0.04}$&	0.04	\\
J0103.2$-$6614	& J010315$-$661425 &	0.33*	&	II	&	186	&	$	1.17^{1.17}_{1.17}$	&	0.28	&	$	0.02^{0.02}_{0.02}$&	0.08	&	$	-0.04^{-0.04}_{-0.05}$&	0.08	&	$	0^{0}_{0.03}$&	0.03	&	$	0.01^{0.01}_{0.01}$&	0.03	\\
J0103.7$-$6632	& J010344$-$663227 &	0.59	&	II	&	-90	&	$	1.2^{1.2}_{1.2}$	&	0.37	&	$	0.04^{0.05}_{0.03}$&	0.11	&	$	-0.12^{-0.11}_{-0.12}$&	0.1	&	$	0.02^{0.01}_{0.04}$&	0.04	&	$	0.04^{0.05}_{0.04}$&	0.05	\\
\hline
\end{tabular}
\end{table}
\end{landscape}

\begin{table*}
 
\begin{minipage}{126mm}

\caption{Sources used to fit the magnitude-redshift relation. Column 1 and 2 give the RA and Dec (J2000) of each source, Columns 3 and 4 give the same for the optical ID for each radio source. Column 5 and 6 give the redshift and magnitude for each source.}
\label{fit_sample}
\begin{tabular}{@{}ccccccc@{}}
\hline

RA (J2000) & Dec (J2000) & RA (Id) (J2000) & DEC (Id) (J2000)& Redshift & $m_{r}$ \\
\hline
0:22:45.10 	&	 $-$66:53:06.3 	&	 0:22:45.10 	&	 $-$66:53:06.3 	&	  0.234 	&	18.48	\\
0:26:14.33 	&	 $-$66:45:55.1 	&	 0:26:14.41 	&	 $-$66:45:54.9	&	  0.426 	&	20.25	\\
0:26:21.47 	&	 $-$67:13:41.6 	&	 0:26:21.54 	&	 $-$67:13:41.7 	&	  0.249 	&	18.45	\\
0:26:21.47 	&	 $-$67:13:41.6 	&	 0:26:21.54 	&	 $-$67:13:41.7 	&	  0.249 	&	18.45	\\
0:26:28.92 	&	 $-$67:21:49.6 	&	 0:26:28.52 	&	 $-$67:21:48.8 	&	  0.274 	&	18.15	\\
0:26:49.08 	&	 $-$66:31:22.0 	&	 0:26:49.10 	&	 $-$66:31:23.0 	&	  0.323 	&	19.3	\\
0:26:49.18 	&	 $-$66:44:00.8 	&	 0:26:49.09 	&	 $-$66:44:01.1 	&	  0.219 	&	17.96	\\
0:27:15.59 	&	 $-$66:24:18.5 	&	 0:27:15.59 	&	 $-$66:24:18.1 	&	  0.073 	&	15.06	\\
0:27:15.60 	&	 $-$66:24:18.7 	&	 0:27:15.50 	&	 $-$66:24:18.6 	&	  0.074 	&	15.06	\\
0:27:46.63 	&	 $-$67:49:51.9 	&	 0:27:46.64 	&	 $-$67:49:52.5 	&	  0.174 	&	16.52	\\
0:27:56.29 	&	 $-$67:37:53.8 	&	 0:27:56.28 	&	 $-$67:37:53.8 	&	  0.252 	&	17.7	\\
0:28:08.51 	&	 $-$66:14:15.9 	&	 0:28:08.55 	&	 $-$66:14:16.5 	&	  0.272 	&	19.22	\\
0:28:09.84 	&	 $-$66:29:38.8	&	 0:28:09.73 	&	 $-$66:29:39.0 	&	  0.339 	&	18.37	\\
0:28:22.42 	&	 $-$66:53:44.1 	&	 0:28:22.41 	&	 $-$66:53:43.5 	&	  0.190 	&	17.32	\\
0:28:29.45 	&	 $-$67:18:43.6 	&	 0:28:29.47 	&	 $-$67:18:44.3	&	  0.243 	&	17.61	\\
0:28:33.98 	&	 $-$67:21:50.2 	&	 0:28:33.98 	&	 $-$67:21:50.3 	&	  0.241 	&	17.83	\\
0:28:41.04 	&	 $-$66:43:45.1 	&	 0:28:41.19 	&	 $-$66:43:44.5 	&	  0.234 	&	18.62	\\
0:28:51.94 	&	 $-$67:58:39.3 	&	 0:28:52.02 	&	 $-$67:58:39.9 	&	  0.352 	&	18.05	\\
0:29:02.47 	&	 $-$66:39:51.6 	&	 0:29:02.65 	&	 $-$66:39:51.9 	&	  0.219 	&	18.57	\\
0:29:04.64 	&	 $-$66:03:20.4 	&	 0:29:04.58 	&	 $-$66:03:21.4 	&	  0.400 	&	19.69	\\
0:29:07.17 	&	 $-$67:22:56.4 	&	 0:29:07.12 	&	 $-$67:22:55.6 	&	  0.220 	&	17.76	\\
0:29:25.65 	&	 $-$67:21:30.7 	&	 0:29:25.55 	&	 $-$67:21:31.5 	&	  0.292 	&	19.38	\\
0:29:44.03 	&	 $-$66:56:23.4 	&	 0:29:43.95 	&	 $-$66:56:23.3 	&	  0.402 	&	19.09	\\
0:29:52.63 	&	 $-$66:06:53.2 	&	 0:29:52.98 	&	 $-$66:06:53.5 	&	  0.402 	&	21.18	\\
0:30:01.70 	&	 $-$67:14:02.2 	&	 0:30:01.69 	&	 $-$67:14:03.3 	&	  0.413 	&	18.63	\\
0:30:09.02 	&	 $-$67:26:44.9 	&	 0:30:09.18 	&	 $-$67:26:45.1 	&	  0.713 	&	21.54	\\
0:30:44.21 	&	 $-$67:36:10.6 	&	 0:30:44.32 	&	 $-$67:36:11.1 	&	  0.321 	&	19.5	\\
0:31:14.79 	&	 $-$67:18:02.0 	&	 0:31:14.68 	&	 $-$67:18:01.5 	&	  0.501 	&	20.64	\\
0:31:17.26 	&	 $-$67:50:52.3 	&	 0:31:17.12 	&	 $-$67:50:53.2 	&	  0.375 	&	19.86	\\
0:31:29.22 	&	 $-$66:55:16.9 	&	 0:31:29.37 	&	 $-$66:55:16.8 	&	  0.532 	&	20.36	\\
0:31:32.05 	&	 $-$67:48:58.7 	&	 0:31:32.49 	&	 $-$67:49:01.1 	&	  0.355 	&	19.1	\\
0:31:32.48 	&	 $-$67:49:00.4 	&	 0:31:32.50 	&	 $-$67:49:00.5 	&	  0.356 	&	19.1	\\
0:31:47.04 	&	 $-$66:20:50.6 	&	 0:31:47.05 	&	 $-$66:20:50.4 	&	  0.278 	&	19.09	\\
0:31:55.80 	&	 $-$66:44:05.8 	&	 0:31:55.81 	&	 $-$66:44:05.1 	&	  0.653 	&	20.95	\\
0:32:01.00 	&	 $-$66:44:06.7 	&	 0:32:00.86 	&	 $-$66:44:06.4 	&	  0.611 	&	20.76	\\
0:32:45.72 	&	 $-$66:29:12.1 	&	 0:32:45.69 	&	 $-$66:29:11.9 	&	  0.214 	&	18.04	\\
0:33:29.46 	&	 $-$67:14:20.2 	&	 0:33:29.35 	&	 $-$67:14:19.2 	&	  0.407 	&	18.38	\\
0:33:46.81 	&	 $-$67:38:03.3 	&	 0:33:46.69 	&	 $-$67:38:04.7 	&	  0.356 	&	18.73	\\
0:33:47.32 	&	 $-$68:00:50.4 	&	 0:33:47.48 	&	 $-$68:00:49.8 	&	  0.225 	&	18.68	\\
0:33:56.56 	&	 $-$66:52:05.7 	&	 0:33:56.48 	&	 $-$66:52:05.4 	&	  0.402 	&	19.1	\\
0:34:05.59 	&	 $-$66:39:34.5 	&	 0:34:05.59 	&	 $-$66:39:34.5 	&	  0.110 	&	16.79	\\
0:34:08.85 	&	 $-$66:26:21.7 	&	 0:34:08.98 	&	 $-$66:26:21.7 	&	  0.486 	&	19.53	\\
0:34:29.18 	&	 $-$66:45:35.7 	&	 0:34:29.19 	&	 $-$66:45:35.6	&	  0.403 	&	19.61	\\
0:34:33.18 	&	 $-$67:36:26.8 	&	 0:34:33.12 	&	 $-$67:36:28.4 	&	  0.069 	&	14.9	\\
0:34:57.51 	&	 $-$66:30:29.8 	&	 0:34:57.44 	&	 $-$66:30:29.6 	&	  0.487 	&	20.27	\\
0:35:02.08 	&	 $-$66:12:52.2 	&	 0:35:01.97 	&	 $-$66:12:52.5 	&	  0.465 	&	19.1	\\
0:35:05.24 	&	 $-$67:41:14.4 	&	 0:35:05.08 	&	 $-$67:41:14.5	&	  0.072 	&	15.32	\\
0:35:34.52 	&	 $-$66:07:24.4 	&	 0:35:34.55 	&	 $-$66:07:25.6 	&	  0.264 	&	19.65	\\
0:35:35.41 	&	 $-$66:56:20.7 	&	 0:35:35.33 	&	 $-$66:56:20.0 	&	  0.296 	&	19.65	\\
0:35:35.85 	&	 $-$66:18:44.3	&	 0:35:35.80 	&	 $-$66:18:44.3	&	  0.508 	&	20.41	\\
0:36:58.16 	&	 $-$66:34:16.3 	&	 0:36:58.17 	&	 $-$66:34:16.4	&	  0.241 	&	18.46	\\
0:36:58.16 	&	 $-$66:34:16.3 	&	 0:36:58.17 	&	 $-$66:34:16.4 	&	  0.241 	&	18.46	\\
\hline
\end{tabular}
\end{minipage}
\end{table*}

\setcounter{table}{2}
\begin{table*}
 
\begin{minipage}{126mm}

\caption{Table.~\ref{fit_sample} continued.}
\label{fit_sample}
\begin{tabular}{@{}ccccccc@{}}
\hline

RA (J2000) & Dec (J2000) & RA (Id) (J2000) & DEC (Id) (J2000)& Redshift & $m_{r}$ \\
\hline
0:37:29.07 	&	 $-$67:02:50.8 	&	 0:37:29.06 	&	 $-$67:02:50.3 	&	  0.350 	&	18.98	\\
0:39:01.31 	&	 $-$67:49:43.7 	&	 0:39:01.39 	&	 $-$67:49:43.5 	&	  0.073 	&	15.08	\\
0:39:03.72 	&	 $-$66:54:36.6 	&	 0:39:03.69 	&	 $-$66:54:34.7 	&	  0.256 	&	18.66	\\
0:40:44.35 	&	 $-$67:24:32.4 	&	 0:40:46.67 	&	 $-$67:24:35.8 	&	  0.296 	&	19.07	\\
0:40:55.50 	&	 $-$66:50:16.1 	&	 0:40:55.46 	&	 $-$66:50:16.5 	&	  0.747 	&	20.86	\\
0:41:00.98 	&	 $-$67:24:32.3 	&	 0:41:01.05 	&	 $-$67:24:33.0	&	  0.299 	&	18.42	\\
0:41:12.20 	&	 $-$67:51:22.2 	&	 0:41:12.33 	&	 $-$67:51:22.1	&	  0.359 	&	20.2	\\
0:41:20.87 	&	 $-$67:08:06.8 	&	 0:41:20.92 	&	 $-$67:08:05.2 	&	  0.492 	&	20.05	\\
0:41:46.39 	&	 $-$67:26:27.5 	&	 0:41:47.29 	&	 $-$67:26:26.8 	&	  0.293 	&	17.34	\\
0:41:46.80 	&	 $-$67:26:15.4 	&	 0:41:46.80 	&	 $-$67:26:15.4 	&	  0.293 	&	19.24	\\
0:41:58.21 	&	 $-$66:54:11.6 	&	 0:41:58.46 	&	 $-$66:54:11.0 	&	  0.520 	&	19.83	\\
0:42:01.69 	&	 $-$67:29:00.8	&	 0:42:01.83 	&	 $-$67:29:03.0 	&	  0.296 	&	18.87	\\
0:42:14.21 	&	 $-$66:54:49.4	&	 0:42:14.25 	&	 $-$66:54:48.6	&	  0.161 	&	18.63	\\
0:42:23.49 	&	 $-$66:25:27.9 	&	 0:42:23.51 	&	 $-$66:25:28.1 	&	  0.210 	&	17.7	\\
0:43:08.54 	&	 $-$66:35:33.2 	&	 0:43:08.59 	&	 $-$66:35:33.9 	&	  0.318 	&	19.23	\\
0:52:06.50 	&	 $-$66:22:51.9 	&	 0:52:07.18 	&	 $-$66:22:55.8 	&	 0.704 	&	21.89	\\
0:56:57.22 	&	 $-$66:32:38.8 	&	 0:56:57.22 	&	 $-$66:32:38.8	&	  0.249 	&	18.6	\\
0:57:04.57 	&	 $-$67:34:11.8 	&	 0:57:04.41 	&	 $-$67:34:12.8 	&	  0.307 	&	19.02	\\
0:57:07.00 	&	 $-$66:32:41.4 	&	 0:57:07.00 	&	 $-$66:32:41.4	&	  0.248 	&	18.81	\\
0:57:12.34 	&	 $-$66:51:17.5 	&	 0:57:06.98 	&	 $-$66:50:59.0 	&	  0.236 	&	18.2	\\
0:57:27.20 	&	 $-$67:03:20.9 	&	 0:57:27.23 	&	 $-$67:03:18.9 	&	  0.260 	&	18.64	\\
0:57:43.62 	&	 $-$67:01:36.2 	&	 0:57:43.62 	&	 $-$67:01:36.2	&	  0.261 	&	18.13	\\
0:57:48.60 	&	 $-$67:02:25.1 	&	 0:57:48.60 	&	 $-$67:02:25.1 	&	  0.064 	&	14.72	\\
1:02:41.48 	&	 $-$67:34:03.1 	&	 1:02:41.48 	&	 $-$67:34:03.1 	&	  0.065 	&	15.53	\\
1:03:10.45 	&	 $-$66:32:21.2	&	 1:03:09.93 	&	 $-$66:32:21.1	&	  0.398 	&	18.87	\\
1:03:14.16 	&	 $-$66:14:40.0 	&	 1:03:14.97 	&	 $-$66:14:24.9 	&	  0.331 	&	18.22	\\
1:03:44.44 	&	 $-$67:47:52.4 	&	 1:03:44.58 	&	 $-$67:47:52.0 	&	  0.329 	&	18.44	\\
1:06:01.66 	&	 $-$66:53:37.0 	&	 1:06:01.85 	&	 $-$66:53:37.1 	&	  0.262 	&	17.6	\\

\hline
\end{tabular}
\end{minipage}
\end{table*}

\label{lastpage}
\end{document}